\documentclass{jfm}
\usepackage{graphicx}
\usepackage{newtxtext}
\usepackage{newtxmath}
\usepackage{natbib}
\usepackage{hyperref}
\hypersetup{
    colorlinks = true,
    urlcolor   = blue,
    citecolor  = black,
}

\newcommand{\RomanNumeralCaps}[1]
\nolinenumbers

\usepackage{xcolor}

\DeclareMathOperator{\sech}{sech}
\DeclareMathOperator{\csch}{csch}

\title{The Onset Acceleration for Surfactant Covered Faraday Waves}

\author{Stephen L. Strickland\aff{1}, Karen E. Daniels\aff{2}, and Michael Shearer\aff{3}}

\affiliation{
\aff{1}Dept. of Physics, Samford University, Homewood, AL, USA 
\aff{2}Dept. of Physics, NC State University, Raleigh, NC, USA 
\aff{3}Dept. of Mathematics, NC State University, Raleigh, NC, USA 
}

\begin{document}
\maketitle

\begin{abstract}
Faraday waves are gravity-capillary waves that emerge on the surface of a vertically vibrated fluid when the energy injected via vibration exceeds the energy lost due to viscous dissipation.
Because this dissipation primarily occurs in the free surface boundary layer, their emergence is particularly sensitive to free surface properties including the surface tension, elasticity, and viscosity of surfactants present at the free surface.
We study this sensitivity by considering a Newtonian fluid bath covered by an insoluble surfactant subject to vertical vibrations which produce sub-harmonic Faraday waves.
By assuming a finite-depth, infinite-breadth, low-viscosity bulk fluid and accounting for surface tension, Marangoni, and Boussinesq effects, we derive an expression for the onset acceleration up to second order in the expansion parameter $\Upsilon = \sqrt{\tfrac{1}{\mathcal{R}e}}$.
We recover the results of previous numerical investigations, but only by modifying the Marangoni and Boussinesq numbers to account for the low-viscosity limit.
The analytic expression allows us to consider a range of parameters not previously studied, including a wide variety of fluid depths and driving frequencies. In addition, we uncover regions of parameter space for which our model predicts that the addition of surfactant would lower, rather than elevate, the onset acceleration.
We discuss the possible use of this model in developing a surface viscometer for surfactant monolayers.
\end{abstract}



\section{Introduction}
\label{s:Intro}

When faced with a roaring ocean, Roman sailors would break open casks of oil, spilling the contents into the sea, and would ride in a patch of quiet oil-covered water until the storm abated.
This calming effect of an oil layer on ocean waves (gravity-capillary waves) has been reported by Pliney the elder \citep{Elder}, publicized by Benjamin Franklin \cite{Franklin1774}, and utilized by Shields at Aberdeen Harbor \cite{Aitken1882} as a way of keeping sea-faring craft safe.
More recently, the dynamic effects of surface materials on gravity-capillary waves have become useful for the 
	remote detection of crude oil spills \cite{Cini1983,Brekke2005,Ghanmi2015},
	detection of biological molecules \cite{Picard2007,Picard2009},
	measurement of bulk and interfacial rheology \cite{Lucassen-Reynders1970,Douady1990,Jiang1993,Raynal1999,Saylor2000,Behroozi2007,Shao2018,Lau2020},
	and the 
	patterning of interfaces \cite{Henderson1991,Henderson1998,Wright-2003-PPF}.

In these applications, a layer of surfactant reduces the surface tension ($\sigma$) of the bulk fluid by an amount that depends upon the surface density ($\Gamma$) of the surfactant, and typically the surface tension decreases monotonically as surfactant density increases.
An inhomogeneous distribution of surfactant will result in surface tension gradients (Marangoni stresses) which drive flow in the bulk fluid.
This flow then transports the surfactant molecules, modifying their spatial distribution.
If left unperturbed, the coupled surfactant-fluid system would reach an equilibrium for which the surfactant becomes homogeneously distributed and the Marangoni stresses vanish via diffusion.

A traveling gravity-capillary wave will compress and expand the surfactant-covered interface, giving rise to Marangoni stresses \citep{Lange1984}.
These stresses in turn result in a viscous boundary layer at the fluid surface where dissipation of the wave's energy is enhanced.
The energy dissipation is often made apparent through the exponential decay of the wave's amplitude as it propagates \citep{Reynolds1880,Levich1941,Dorrestein1951,Case1956,Goodrich1961a,Davies1965,Lucassen-Reynders1970,Jiang1993,Saylor2000,Behroozi2007}.

For linear small-amplitude gravity-capillary waves, the energy dissipation rate is characterized by the damping parameter $\delta$ and is related to the surface elastic modulus (a.k.a. surface dilational modulus or Gibbs' elasticity), $\varepsilon_0 = - \Gamma_0 \tfrac{d \sigma}{d \Gamma}$ where $\Gamma_0$ is the equilibrium mean surfactant density.
Contemporary theoretical and experimental research \citep{Reynolds1880,Levich1941,Dorrestein1951,Case1956,Goodrich1961a,Davies1965,Miles1967,Lucassen-Reynders1970,Jiang1993,Saylor2000,Behroozi2007} has shown that the damping $\delta$ increases non-monotonically as a function of $\varepsilon_0$.
For $\varepsilon_0 = 0$, $\delta$ measures the bulk damping effect in the absence of surfactant.
As $\varepsilon_0$ is increased (typically by adding more surfactant), $\delta$ reaches a maximum that can be an order of magnitude larger than the surfactant-free bulk damping.
For larger $\varepsilon_0$, $\delta$ decreases to a value that is roughly half of its peak value \citep{Reynolds1880,Levich1941,Dorrestein1951,Case1956,Goodrich1961a,Davies1965,Miles1967,Lucassen-Reynders1970,Jiang1993,Saylor2000,Behroozi2007}.

Small-amplitude standing gravity-capillary waves, known as Faraday waves \citep{Faraday1831a}, emerge when a fluid bath is vertically vibrated above an angular frequency $\omega_c$, provided the acceleration amplitude $a$ meets or exceeds the onset acceleration $a_{c}$ for that frequency.
The emergent wave, with angular frequency $\omega_0$, can either be harmonic ($\omega_0 = \omega$) or sub-harmonic ($\omega_0 = \tfrac{1}{2} \omega$).
In this work, we focus exclusively on sub-harmonic Faraday waves.

We understand the emergence of the Faraday waves from an energy-balance standpoint: the fluid bath dissipates energy in every wave mode due to the viscosity of the fluid.
On the other hand, vertical vibration injects energy into all wave modes, but not uniformly.
When the driving amplitude is less than the onset acceleration, the energy dissipation exceeds energy injection in all modes so that no wave emerges, but at the onset acceleration $a_{c}$, a single wave mode has more energy injected than dissipated, while all others remain dissipative. Therefore,  a wave emerges with a selected pattern of wavenumber $k_{c}$ \citep{Edwards1994,Gollub2006,Ibrahim2015}.

Because the energy dissipation largely occurs in the boundary layer near the free surface, the onset acceleration for Faraday waves is very sensitive to the presence of surface stresses such as due to surface tension, surface elasticity, and surface viscosity.
Because surfactants modify these stresses, the onset of Faraday waves can serve as an effective indicator of the presence of a surfactant as well as a means of measuring the rheological properties of the surfactant layer.
The effects of soluble surfactant on the Faraday wave onset have been observed \citep{Ballesta2005} as has the effect of insoluble surfactant on the damping rates of Faraday waves \citep{Henderson1991,Henderson1998}.

\citet{Benjamin1954,Kumar1994,chen1999}, and \citet{Kumar2002,Kumar2004,Kumar2004a} have made theoretical predictions for these non-linear waves that relate the onset acceleration to the rate of energy dissipation in the system.
These theoretical predictions for the onset acceleration are typically formulated by using Floquet analysis, first applied by \citet{Kumar1994}, which results in a recursion relation whose truncation is often solved with numerical techniques for a pre-specified parameter regime.
This combination of Floquet analysis and numerical solvers has been expanded to consider surfactant effects by \citet{Ubal2005,Ubal2005a,Ubal2005b,Giavedoni2007,Kumar2004a,Mikishev2016}, who found that the onset acceleration is sensitive to the rheological properties of the surfactant layer in much the same way that the viscous damping parameter $\delta$ for linear gravity-capillary waves depends upon the surface elasticity $\varepsilon_0$.
Using a  different approach (a purely analytic technique), \citet{chen1999} considered Faraday waves on an infinite-depth surfactant-free fluid and derived an exact expression for the onset acceleration and wave number of the emergent Faraday waves.
Chen and Vi\~{n}als also started with Floquet analysis yielding a recursion relation, but instead of solving this relation numerically, they considered the weak viscosity limit, expanding the driving acceleration and wave number in terms of $\gamma = \tfrac{1}{\mathcal{R}e}$ and solving for the coefficients of the expansion.

In this paper, we extend the techniques of \citet{chen1999} into the finite-depth low-viscosity regime with surfactant, and we show that our analysis improves upon the numerical predictions of \citet{Kumar2004} and \citet{Giavedoni2007}.
In section~\S\ref{s:GovernEqns}, we present the parameterization of our system and the linearized governing equations for our model.
The techniques for solving these equations are in section~\S\ref{s:Solving} while the general solution and special cases are in section~\S\ref{s:Solutions}.
In section~\S\ref{s:GiavedoniTest}, we compare our analytic solution to the results of previous numerical investigations.
Novel features of the onset acceleration are in section~\S\ref{s:ExploreAnalyticalSoln}, and a possible application of this model for developing a surface viscometer is detailed in section~\S\ref{s:Conclusion}.

\section{Parameterization \& Governing Equations}
\label{s:GovernEqns}

\begin{figure}
\centering
\includegraphics[width=0.70\linewidth]{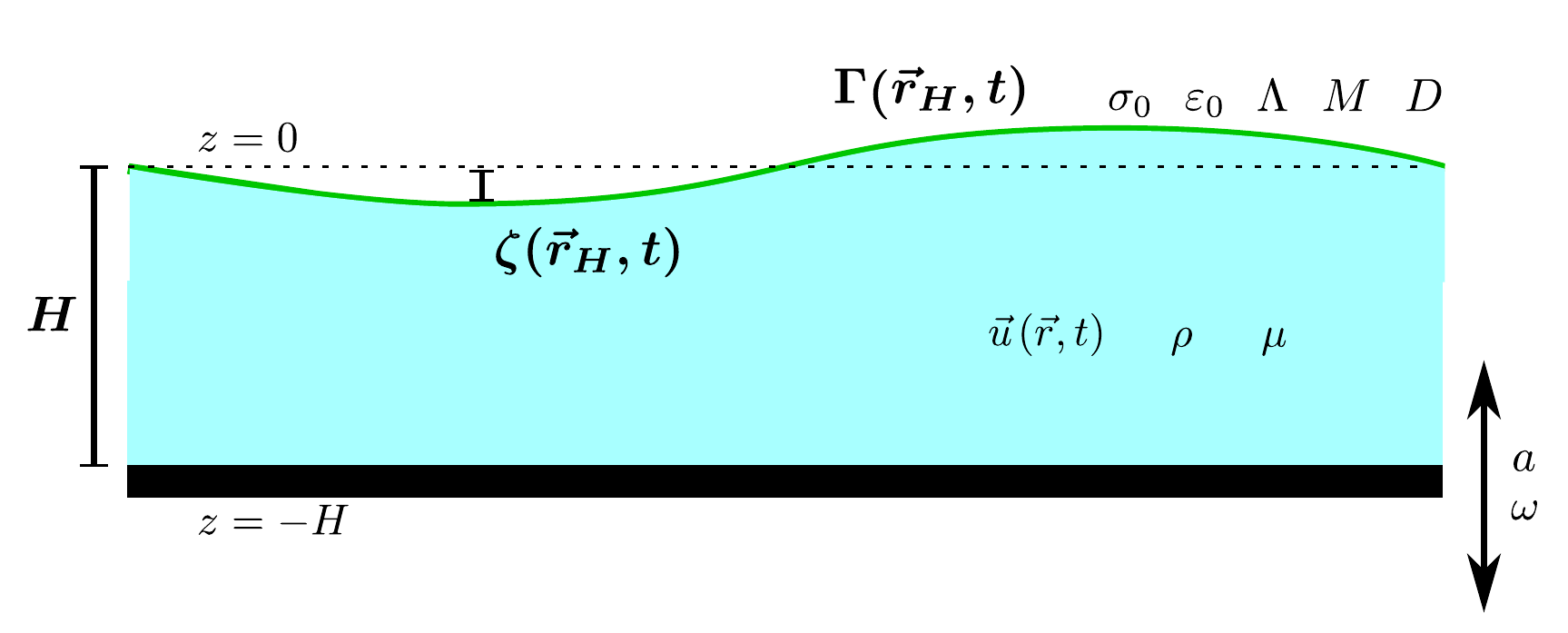}
\caption{
	Schematic showing a horizontally-infinite layer of incompressible fluid with mean finite depth $H$ covered with a surfactant layer (green) subject to vertical vibrations of angular frequency $\omega$ with an amplitude $a$ driven by the container floor (black).
	The flat equilibrium fluid surface (dashed line) is taken as $z=0$ and the perturbation of that surface is $\zeta$.
	The bulk fluid is assumed to have  constant density $\rho$, dynamic viscosity $\mu$, and flow velocity $\vec{u}$.
	The free surface have spatiotemporally-varying surfactant mass density $\Gamma$, equilibrium surface tension $\sigma_0$, surface elasticity $\varepsilon_0$, surface dilational viscosity $\Lambda$, surface shear viscosity $M$, and surface diffusivity $D$.
}
\label{f:ThrySchematic}
\end{figure}

\begin{table}
	\centering
	\begin{tabular}{ c | c }
		Symbol & Description \\[0.2ex]
		\hline \hline
		$\left[x,y,z\right]$ & Cartesian coordinates \\[0.2ex]
		$\vec{r}_H$ & horizontal coordinates \\[0.2ex]
		$H$ & bulk fluid depth \\[0.2ex]
		$\zeta$ & surface elevation \\[0.2ex]
		$\vec{u} = \left[ u , v , w \right]$ & flow velocity \\[0.2ex]
		$\rho$ & bulk fluid density \\[0.2ex]
		$\mu$ & bulk fluid dynamic viscosity \\[0.2ex]
		$\delta$ & damping parameter \\[0.2ex]
		$\Gamma$ , $\Gamma_0$ & surfactant surface density  \& mean equilibrium surface density \\[0.2ex]
		$\sigma$ , $\sigma_0$ & surface tension \& mean equilibrium surface tension \\[0.2ex]
		$\varepsilon_0$ & surface elastic modulus \\[0.2ex]
		$\Lambda$ & surface dilational viscosity \\[0.2ex]
		$M$ & surface shear viscosity \\[0.2ex]
		$\Omega = \Lambda + 2 M$ & combined surface viscosity \\[0.2ex]
		$D$ & surface diffusivity \\[0.2ex]
		$g$ & gravitational constant \\[0.2ex]
		$a$ & driving acceleration amplitude \\[0.2ex]
		$\omega$ & driving frequency \\[0.2ex]
		$\omega_0 = \tfrac{1}{2} \omega$ & sub-harmonic Faraday wave frequency \\[0.2ex]
		$a_c$ & critical (onset) acceleration \\[0.2ex]
		$k_c$ & critical (onset) wavenumber \\[0.2ex]
		$k_0$ & wavenumber for gravity-capillary waves on an unvibrated fluid \\[0.2ex]
		$l_0 = 1/k_0$ & lengthscale \\[0.2ex]
	\end{tabular}
	\caption{
		Physical parameters
	}
	\label{t:Param}
\end{table}

\begin{table}
	\centering
	\begin{tabular}{ c | c }
		Definition & Description \\[2ex]
		\hline \hline
		$ \mathcal{F} = \frac{\displaystyle a}{\displaystyle g} $ & dimensionless acceleration amplitude \\[2ex]
		$ G = \frac{\displaystyle g}{\displaystyle l_0 \omega_0^2}$ & dimensionless gravitational acceleration \\[2ex]
		$\Sigma = \frac{\displaystyle \sigma_0}{\displaystyle \rho \omega_0^2 l_0^3}$ & dimensionless surface tension \\[2ex]
		$\frac{\displaystyle G}{\displaystyle \Sigma} = \frac{\displaystyle \rho g l_0^2}{\displaystyle \sigma_0} $ & Bond number \\[2ex]
		$\mathcal{R}e = \frac{\displaystyle \rho \omega_0 l_0^2}{\displaystyle \mu}$ & bulk Reynolds number \\[2ex]
		$ \Upsilon = \sqrt{\frac{\displaystyle \mu}{\displaystyle \rho \omega_0 l_0^2}} = \sqrt{\frac{\displaystyle 1}{\displaystyle \mathcal{R}e}} $ & the expansion parameter \\[2ex]
		$\mathcal{C}a = \frac{\displaystyle 1}{\displaystyle \mathcal{R}e \Sigma} = \frac{\displaystyle \mu \omega_0 l_0}{\displaystyle \sigma_0} $ & capillary number \\[2ex]
		$\mathcal{M} = \frac{\displaystyle \varepsilon_0/\sigma_0}{\displaystyle \mathcal{C}a} = \frac{\displaystyle \varepsilon_0}{\displaystyle \mu \omega_0 l_0}$ & Marangoni number \\[2ex]
		$ \mathcal{M}^{\dagger} = \Upsilon \mathcal{M} = \frac{\displaystyle \varepsilon_0 }{\displaystyle \sqrt{ \rho \mu \omega_0^3 l_0^4} } $ & modified Marangoni number \\[2ex]
		$\mathcal{P}e = \frac{\displaystyle \omega_0 l_0^2}{\displaystyle D}$ & surface Peclet number \\[2ex]
		$\mathcal{B}_D = \frac{\displaystyle \Lambda + M}{\displaystyle \mu l_0}$ & dilational Boussinesq number \\[2ex]
		$\mathcal{B}_S = \frac{\displaystyle M}{\displaystyle \mu l_0}$ & shear Boussinesq number \\[2ex]
		$\mathcal{B} = \mathcal{B}_D + \mathcal{B}_S = \frac{\displaystyle \Omega}{\displaystyle \mu l_0}$ & combined Boussinesq number \\[2ex]
		$ \mathcal{B}^{\dagger} = \Upsilon \mathcal{B} = \frac{\displaystyle \Omega}{\displaystyle \sqrt{\rho \mu \omega_0 l_0^4} } $ & modified Boussinesq number \\[2ex]
	\end{tabular}
	\caption{
		Dimensionless numbers
	}
	\label{t:Non-dim}
\end{table}

We consider, as shown in figure \ref{f:ThrySchematic} and parameterized in table \ref{t:Param}, an incompressible Newtonian bulk fluid  of infinite horizontal extent and finite depth $H$ with flow velocity $\vec{u}(\vec{r},t) = [ u , v , w ]$, density $\rho$, and dynamic viscosity $\mu$.
The elevation of the free surface is given by $z = \zeta(\vec{r}_H,t)$ where we denote the horizontal coordinates as $\vec{r}_H$.
We will use the subscript $_H$ to indicate a projection   onto the horizontal plane. The dynamics of the air above the free surface are taken to be negligible.

The surfactant monolayer at the free surface is treated as a 2-dimensional Newtonian fluid.
At equilibrium the mean surfactant mass density is uniformly $\Gamma_0$ which determines the mean surface tension $\sigma_0(\Gamma_0)$, mean surface dilational viscosity $\Lambda(\Gamma_0)$, mean surface shear viscosity $M(\Gamma_0)$, and mean surface diffusivity $D(\Gamma_0)$.
When the surface is dynamic, the instantaneous local surfactant mass density is $\Gamma(\vec{r}_H,t)$ thereby inducing small negligible perturbations in $\Lambda$, $M$, and $D$ away from their equilibrium values.
Variations in the surface tension $\sigma = \sigma(\Gamma)$ are significant however; in linearizing the governing equations, effects of surface tension are divided into the equilibrium surface tension $\sigma_0$ and the surface elasticity $\varepsilon_0 = - \Gamma \tfrac{d \sigma}{d \Gamma}$ which will both be considered constants.

To ensure that the unperturbed fluid surface remains at $z=0$ during vibration, we will analyze the system in the non-inertial reference frame that is co-vibrated with the container floor.
In effect, this gives a gravitational body-force term $\vec{g}(t) = - \left( g + a \cos(\omega t) \right) \hat{z}$ in which $g=9.8$~m/s$^2$ is the gravitational constant and $a$ is the acceleration amplitude of the vibration.

In developing the model, we non-dimensionalize the governing equations by choosing scales for time, length, and mass.
We scale time by $\tfrac{1}{\omega_0}$, the Faraday wave period.
The length scale is set by $l_0 = \tfrac{1}{k_0}$, where $k_0$ is the wavenumber for a gravity-capillary wave on an unvibrated fluid as given by the finite-depth Kelvin dispersion relation $ 1 = \left( \tfrac{g}{\omega_0^2} k_0 + \tfrac{\sigma_0}{\rho \omega_0^2} k_0^3 \right) \tanh( k_0 H ) $.
We scale mass by $\rho l_0^3$.
Velocities are scaled by $l_0 \omega$, and $\zeta$ and $H$ are scaled by a factor of $l_0$.
We also scale the surfactant density by $\Gamma_0$ and pressure by $\rho l_0^2 \omega_0^2$.
This process results in the standard dimensionless numbers defined in table \ref{t:Non-dim}.
For the remainder of this manuscript, $\vec{u}$, $p$, $k$, $H$, $\zeta$, and $\Gamma$ will refer to the dimensionless versions of these quantities.

To analyze the system at onset, we linearize the governing equations and boundary conditions around the equilibrium $\vec{u} = 0$, $\zeta = 0$, $\Gamma = 1,$ which  refer to as the trivial solution.
The bulk fluid satisfies the linearized incompressible Navier-Stokes equations,
\begin{equation}
	\label{eqn:SubNS}
	\partial_t \vec{u} = - \vec{\nabla} p + \frac{1}{\mathcal{R}e} \nabla^2 \vec{u} - G \left( 1 + \mathcal{F} \cos( 2 t ) \right) \hat{z}
\end{equation}
\begin{equation}
	\label{eqn:SubCont}
	  \vec{\nabla} \cdot \vec{u} =0
\end{equation}
with Reynolds number $\mathcal{R}e$, gravitation number $G$, and dimensionless driving acceleration $\mathcal{F}$.
The bulk fluid also satisfies the no-slip boundary condition at the floor $z = -H:$  
\begin{equation}
	\label{eqn:BCBot}
	  \vec{u} = 0, \ \ \  z = -H.
\end{equation}
At the free surface $z = \zeta(\vec{r}_H,t)$, we have further boundary conditions: the kinematic boundary condition,
\begin{equation}
	\label{eqn:BCKin}
	\partial_t \zeta = w,
\end{equation}
and the surface continuity equation, expressing the advection and diffusion of the surfactant,
\begin{equation}
	\label{eqn:SurfCont}
	0 = \partial_t \Gamma + \vec{\nabla}_H \cdot \vec{u}_H - \frac{1}{\mathcal{P}e} \nabla_S^2 \Gamma
\end{equation}
with Peclet number $\mathcal{P}e$.

Since the mass of the surfactant monolayer is negligible, the surface tangential and normal stress boundary conditions contain no inertial or gravitational terms for the surfactant,
\begin{equation}
	\label{eqn:SurfStressTan}
	0 = - \left[ \begin{array}{c} \partial_x w + \partial_z u \\ \partial_y w + \partial_z v \end{array} \right] - \mathcal{M} \vec{\nabla}_H \Gamma + \mathcal{B}_D \vec{\nabla}_H \vec{\nabla}_H \cdot \vec{u}_H + \mathcal{B}_S \vec{\nabla}_H^2 \vec{u}_H,
\end{equation}
\begin{equation}
	\label{eqn:SurfStressNorm}
	0 = \frac{1}{\Sigma} p - 2 \mathcal{C}a \, \partial_z w - \nabla_H^2 \zeta
\end{equation}
with Marangoni number $\mathcal{M}$ and dilational and shear Boussinesq numbers $\mathcal{B}_D$ and $\mathcal{B}_S$ respectively.
General governing equations for a surfactant-covered surface are given by  \citet{Scriven1960}, with later corrections by \citet{Waxman1984}.

The equations can be reduced and expressed in terms of $w$, $\zeta$, and $\Gamma$ exclusively.
In the reduction, the dilational and shear Boussinesq effects can be combined, so it is convenient to define an effective surface viscosity  $\Omega = \Lambda + 2 M$ and an effective Boussinesq number   $\mathcal{B} = \mathcal{B}_D + \mathcal{B}_S$.
The equations and boundary conditions are then

\begin{subequations}
	\label{eqn:LinEqs}
	\begin{equation}
		\label{eqn:LinEq-NSzModD} 
		0 = \left[ \partial_t ( \nabla_H^2 + \partial_{zz} ) - \frac{1}{\mathcal{R}e} ( \nabla_H^2 + \partial_{zz} )^2 \right] w
	\end{equation} 
	\begin{equation}
		\label{eqn:LinEq-BotBC1} 
		 w=0;  \ \ \  
		  \partial_z w =0, \ \
		  {z = -H}
	\end{equation} 
	\begin{equation}
		\label{eqn:LinEq-Kin} 
		\partial_t \zeta = w, \ \ \   
		{z = \zeta}
	\end{equation} 
	\begin{equation}
	\label{eqn:LinEq-Cont} 
		\partial_t \Gamma - \partial_z w - \frac{1}{\mathcal{P}e} \nabla_H^2 \Gamma =0, \ \ \ 
		{z = \zeta}
	\end{equation} 
	\begin{equation}
		\label{eqn:LinEq-Tan} 
		- \left[ \nabla_H^2 - \partial_{zz} \right] w 
		- \mathcal{M} \nabla_H^2 \Gamma 
		- \mathcal{B} \nabla_H^2 \partial_z w =0, \ \ \ 
		{z = \zeta}
	\end{equation} 
	\begin{equation}
		\label{eqn:LinEq-Norm} 
				\mathcal{C}a~\partial_z \left( \mathcal{R}e \partial_t - \left( 3 \nabla_H^2 + \partial_{zz} \right) \right) w 
				 - \tfrac{G}{\Sigma} \left( 1 + \mathcal{F} \cos(2 t) \right) \nabla_H^2 \zeta 
				+ \nabla_H^2 \nabla_H^2 \zeta =0,
				\ \ \ 
				{z = \zeta}
	\end{equation} 
\end{subequations}
with Capillary number $\mathcal{C}a$, dimensionless surface tension $\Sigma$, and Bond number $\tfrac{G}{\Sigma}$.

\section{Technique for Finding the Onset Acceleration}
\label{s:Solving}

The onset acceleration $\mathcal{F}_{c} = \tfrac{a_{c}}{g}$ is the minimum value of the driving parameter $\mathcal{F}$ for which $w$, $\zeta$, and $\Gamma$ are non-trivial.
As detailed in appendix \ref{A1:Constants}, we apply an ansatz and solve for non-trivial $w$, $\zeta$, and $\Gamma$ up to a family of constants $\zeta_j$ which are the amplitudes of each wave mode.
We will see that these wave mode amplitudes are coupled in a way that allows us to solve for the onset acceleration.

For a given wavenumber $k$, the ansatz becomes:
\begin{equation}
	\label{eqn:Ansatz}
		\begin{aligned}
			w &= \cos( \vec{k} \cdot \vec{r}_H ) \sum_{j \in \mathbb{Z}_{\text{odd}}} i j e^{ i j t } \left( \mathcal{A}_{j} \sinh(k z) + \mathcal{B}_{j} \cosh(k z) + \mathcal{C}_{j} \sinh(q_{j} z) + \mathcal{D}_{j} \cosh(q_{j} z) \right) 
			\\
			\zeta &= \cos( \vec{k} \cdot \vec{r}_H ) \sum_{j \in \mathbb{Z}_{\text{odd}}} \zeta_j e^{ i j t } 
			\\
			\Gamma &= 1 + \cos( \vec{k} \cdot \vec{r}_H ) \sum_{j \in \mathbb{Z}_{\text{odd}}} \Gamma_j e^{ i j t } 
		\end{aligned}
\end{equation}
where $q^2_{j} = k^2 + i j \mathcal{R}e$.
The series for $w$ represents the general solution of the fourth order partial differential equation \eqref{eqn:LinEq-NSzModD}, which is satisfied by any choice of coefficients $\mathcal{A}_{j}$, $\mathcal{B}_{j}$, $\mathcal{C}_{j}$, and $\mathcal{D}_{j}$. 
Equation \eqref{eqn:LinEq-Cont} then gives a homogeneous equation expressing the coefficients $\Gamma_j$ as linear combinations of $\mathcal{A}_{j}$, $\mathcal{B}_{j}$, $\mathcal{C}_{j}$, and $\mathcal{D}_{j}$.
Together, the homogeneous equations \eqref{eqn:LinEq-BotBC1}, \eqref{eqn:LinEq-Kin}, and \eqref{eqn:LinEq-Tan} form a linear system showing that the coefficients $\mathcal{A}_{j}$, $\mathcal{B}_{j}$, $\mathcal{C}_{j}$, and $\mathcal{D}_{j}$ not only depend on $k,$ but also are all proportional to $\zeta_j$. 

We are left with equation \eqref{eqn:LinEq-Norm}, in which the temporal modes are coupled due to the forcing term $\mathcal{F} \cos(2 t)$.
In this term, $\zeta_j e^{i j t} \cos(2t)$ splits into $\zeta_j e^{i (j+2) t}$ and $\zeta_j e^{i (j-2) t}$, so equation \eqref{eqn:LinEq-Norm} becomes a linear difference equation for the sequence  of coefficients $\{\zeta_j\}$:
\begin{equation}
	\label{eqn:RecursRelat}
		0 = - \zeta_j H_j + \zeta_{j-2} \mathcal{F} + \zeta_{j+2} \mathcal{F}
.
\end{equation}
Here, the term $\zeta_j H_j$ is given by the formula:
\begin{equation}
	\label{eqn:HDef}
		\zeta_j H_j = - \frac{2}{G} \left[
			\zeta_j G
			+ \zeta_j \Sigma k^2
			+ \frac{i j}{k \mathcal{R}e} \left(
				\mathcal{A}_{j} ( k^2 + q_{j}^2 ) 
				+ 2 k q_{j} \mathcal{C}_{j}
			\right)
			\right]
.
\end{equation}
Since $\mathcal{A}_{j}, \mathcal{C}_{j}$ are proportional to $\zeta_j$, the right hand side of equation \eqref{eqn:HDef} is also proportional to $\zeta_j$. Consequently, each $H_j$ depends only on the wavenumber $k.$ 

Following \citet{chen1999}, to solve for the onset acceleration $\mathcal{F}_c$ (the minimum $\mathcal{F}$), we truncate this difference equations as follows.
We note that when a wave of wavenumber $k$ is driven near its onset, it oscillates sub-harmonically indicating that $\zeta_1$ is the most significant contributor with nearly all other modes being negligible.
But as the driving exceeds the onset, higher order frequencies emerge making the higher order $\zeta_j$ more significant.
Since we want to solve for the onset, we can consider the $\zeta_j$ to approach $0$ as $j\to \infty$, suggesting that truncation at large $j=n$ will provide increasingly accurate estimates. 
Starting with $\zeta_n$, we use the recursion relation to solve for $\zeta_j, \  j\leq n:$
\begin{align}
	\label{eqn:ResursRelat2}
	\zeta_n =& \zeta_{n-2}  \frac{\mathcal{F}}{H_n} 
	&
	\zeta_{n-2} =& \zeta_{n-4} \frac{\mathcal{F}}{H_{n-2} - \frac{\mathcal{F}^2}{H_{n}}} 
	&
	\ldots
\end{align}
Extending to  $\zeta_{1}$, we obtain:
\begin{equation}
	\label{eqn:ResursRelat3}
	\zeta_1 = \zeta_{-1} \frac{\mathcal{F}}{H_1 - \frac{\mathcal{F}^2}{H_3 - \frac{\mathcal{F}^2}{H_5 - \ldots}}}
\end{equation}
Because $\zeta$ (which measures the displacement of the fluid surface) is real valued, $\zeta_{-1} = \zeta_{1}^*$ and therefore $ \tfrac{\zeta_1}{\zeta_{-1}} = e^{i \phi} $ where $ \phi = 2 \arg(\zeta_1)$ measures the phase difference between the driving vibration (of frequency $\omega$) and the wave oscillation (of frequency $\omega_0 = \tfrac{1}{2} \omega$).
Eliminating the $\zeta_j$, we obtain an expression relating $\mathcal{F}$ and $k:$
\begin{equation}
	\label{eqn:RecurRelatSoln}
	\mathcal{F} = e^{i \phi} \left( H_1 - \frac{ \mathcal{F}^2 }{ H_3 - \frac{\mathcal{F}^2}{ H_5 - \ldots } } \right)
	.
\end{equation}
Since  the coefficients $H_j$  depend only on the wavenumber $k,$ this equation defines the driving acceleration $\mathcal{F}$ implicitly as a function of $k.$ The onset acceleration $\mathcal{F}_{c}=\mathcal{F}(k_c)$ is the minimum of $\mathcal{F}(k),$ with onset wavenumber $k=k_{c}.$


In finding $\mathcal{F}_{c}$, one could proceed with a first derivative test as done by \citet{chen1999}.
Here, we instead consider that, as with a driven damped oscillator, resonance occurs when the driver optimally injects energy into the oscillator.
Often expressed in terms of a phase difference of $\tfrac{\pi}{2}$ between the forcing and the position of a driven oscillator, a similar phase difference happens with Faraday waves near onset.
As shown in Figure~1 of \citet{Douady1989}, the vibration acceleration and the Faraday wave have a phase difference $\phi = \tfrac{\pi}{2}$, and on close examination, the solution reported by \citet{chen1999} exhibits this same phase difference. Consequently, at the onset acceleration $\mathcal{F} = \mathcal{F}_{c}$, we have  $ \tfrac{\zeta_1}{\zeta_{-1}} = e^{i \phi} = i $, and  \eqref{eqn:RecurRelatSoln} becomes:
\begin{equation}
	\label{eqn:RecurRelatSoln2}
	0 = i \mathcal{F}_{c} + H_1 - \frac{ \mathcal{F}_{c}^2 }{ H_3 - \frac{\mathcal{F}_{c}^2}{ H_5 - \ldots } }.
\end{equation}

To solve equation \eqref{eqn:RecurRelatSoln2} in the case of weak bulk viscosity, we express the onset wavenumber $k_{c}$ and onset acceleration $\mathcal{F}_{c}$ as power series in the parameter $\Upsilon = \sqrt{\tfrac{1}{\mathcal{R}e}}$. 
In the limit $\Upsilon \to 0$, we observe $k_{c} \to 1$ and $\mathcal{F}_{c} \to 0$, so that the power series become:
\begin{equation}
	\label{eqn:PowerSeries}
		\mathcal{F}_{c} = \sum_{n=1}^{\infty} \alpha_n \Upsilon^n,
		\ \ \  \ \ \ 
		k_{c} = 1 + \sum_{n=1}^{\infty} \beta_n \Upsilon^n.
\end{equation}

In considering the weak viscosity limit, the behaviors of the Marangoni and Boussinesq numbers need careful attention.
A naive consideration of Table \ref{t:Non-dim} would suggest that $\mathcal{M}\sim\mathcal{B}\sim\tfrac{1}{\mu}\sim\mathcal{O}(\Upsilon^{-2})$; however, in using the method of dominant balance (see appendix \ref{A2:Expansion}), we find $\mathcal{M}\sim\mathcal{B}\sim\Upsilon^{-1}$ and therefore define modified Marangoni and modified Boussinesq numbers $\mathcal{M}^{\dagger}$ and $\mathcal{B}^{\dagger}$ as: 
\begin{equation}
	\label{eqn:ModSurfNumbers}
		\mathcal{M}^{\dagger} = \frac{\varepsilon_0}{\sqrt{\mu \rho \omega_0^3 l_0^4}} 
		\ \ \ , \ \ \ 
		\mathcal{B}^{\dagger} = \frac{\Omega}{\sqrt{\mu \rho \omega_0 l_0^4}}
\end{equation}
so that $\mathcal{M}=\Upsilon^{-1}\mathcal{M}^{\dagger}$ and $\mathcal{B}=\Upsilon^{-1}\mathcal{B}^{\dagger}$ with $\mathcal{M}^{\dagger}\sim\mathcal{B}^{\dagger}\sim\mathcal{O}(1)$.
The particular choice $\mathcal{M} = \Upsilon^{-1} \mathcal{M}^{\dagger}$ provides a way of putting the surfactant effects as $\mathcal{O}(1)$ in $H_j$ rather than $\mathcal{O}(\Upsilon)$.
If one were to take $\mathcal{M} = \mathcal{M}^{\dagger}$, the onset acceleration would monotonically increase without bound as the Marangoni number increases.
Alternatively, if one were to take $\mathcal{M} = \Upsilon^{-2} \mathcal{M}^{\dagger}$, then all of the onset acceleration terms second order and higher would be infinite.
In choosing $\mathcal{M} = \Upsilon^{-1} \mathcal{M}^{\dagger}$, we obtain quantitative agreement with previous numerical work as we will show in section \S\ref{s:GiavedoniTest}.

At this point, we solve for the $\alpha_n$ and $\beta_n$ by substitution into eqn \ref{eqn:RecurRelatSoln2} and collect terms of like order in $\Upsilon$.
We find that the complex valued equations permit real valued solutions for $\alpha_n$ and $\beta_n$.
For reference, appendix \ref{A2:Expansion} shows the $H_j$ expanded in terms of $\Upsilon$.

\section{The solution}
\label{s:Solutions}

Before presenting the general solution for the onset acceleration, it is reassuring to consider the solution in a few specific cases that have already been well-studied.
The original analytical work by \citet{chen1999} considered the infinite-depth surfactant-free problem, so in section \S\ref{ss:NoSurfLim}, we will show that our analysis recovers their result.
In section \S\ref{ss:FinDepth}, we will extend the surfactant-free problem to the finite-depth limit.
In section \S\ref{ss:SurfLim}, we will consider the infinite-depth surfactant problem since the Marangoni, Boussinesq, and Peclet effects are easier to identify.
We will then present the complete solution for a finite-depth surfactant-covered fluid in section \S\ref{ss:GenSoln}.
For all cases, Mathematica was used to help calculate the coefficients of the series expansions.

\subsection{The surfactant-free infinite-depth case}
\label{ss:NoSurfLim}

We obtain the surfactant-free infinite-depth case by letting
	$\mathcal{M}^{\dagger} \rightarrow 0$,
	$\mathcal{B}^{\dagger} \rightarrow 0$,
	and $H \rightarrow \infty$.
In this case, eqn \eqref{eqn:RecurRelatSoln2} becomes:
\begin{equation}
	\begin{aligned} 
		0 =& - \tfrac{2}{G} \left( - 1 + G + \Sigma \right) & & \\ 
			&+ \Upsilon \left( 
				i \alpha_1 
				- \tfrac{2}{G} \beta_1 \left( 1 + 2 \Sigma \right) 
			\right) \\
			&+ \Upsilon^2 \left( 
				i \left( \alpha_2 - \tfrac{8}{G} \right)
				- \tfrac{2}{G} \beta_2 ( 1 + 2 \Sigma )
			\right) \\
			&+ \Upsilon^3 \left( 
				i \left( \alpha_3 + \tfrac{4 \sqrt{2}}{G} \right)
				- \tfrac{2}{G} ( \beta_3 ( 1 + 2 \Sigma ) - 2 \sqrt{2} )
			\right) \\
			&+ \Upsilon^4 \left( 
				i \alpha_4
				- \tfrac{2}{G} ( \beta_4 ( 1 + 2 \Sigma ) + 4 + \tfrac{G^2}{32} \alpha_2^2 )
			\right) \\
			&+ \Upsilon^5 \left( 
				i \left( \alpha_5 - \tfrac{2 \sqrt{2} + 8 \beta_3}{G} \right)
				- \tfrac{2}{G} ( \beta_5 ( 1 + 2 \Sigma ) - \sqrt{2} + \tfrac{G^2}{16} \alpha_2 \alpha_3 )
			\right) \\
			&+ \mathcal{O}( \Upsilon^6 )
	\end{aligned} 
\end{equation}
Because $\Upsilon$ is arbitrary, each term must independently vanish.
The zeroth order term is a reiteration of the infinite-depth Kelvin dispersion relation, and the higher order terms permit solutions for the $\alpha_j$ , $\beta_j$.
The onset acceleration and wavenumber thus become:
\begin{subequations}
	\begin{equation}
		\label{eqn:CVAon}
		\mathcal{F}_{c} = \frac{ 1 }{ G } \left[ 8 \Upsilon^2 - 4 \sqrt{2} \Upsilon^3 + \frac{ 2 \sqrt{2} ( 11 - 2 G ) }{ ( 3 - 2 G ) } \Upsilon^5 + \mathcal{O}( \Upsilon^6 ) \right]
	\end{equation}
	\begin{equation}
		\label{eqn:CVkon}
		k_{c} = \left[ 1 + \frac{ 2 \sqrt{2} }{ 3 - 2 G } \Upsilon^3 - \frac{6}{ 3 - 2 G } \Upsilon^4 + \frac{ 3 \sqrt{2} }{ 3 - 2 G } \Upsilon^5 + \mathcal{O}( \Upsilon^6 ) \right]
	\end{equation}
\end{subequations}
The expression for the onset acceleration $\mathcal{F}_{c}$ is identical to \citet{chen1999}, where $ \Upsilon = \sqrt{ \gamma / 2 } $ in their notation.
However, their expression for the wavenumber $k_{c}$ is the same only up to third order, so the formula in \citet{chen1999} is not a valid solution to equation \eqref{eqn:RecurRelatSoln2}.

\subsection{The surfactant-free finite-depth case}
\label{ss:FinDepth}

The surfactant-free finite-depth case is achieved by letting $\mathcal{M}^{\dagger} \rightarrow 0$ and $\mathcal{B}^{\dagger} \rightarrow 0$ while keeping $H$ arbitrary.
The onset acceleration becomes: 
\begin{equation}
	\label{eqn:FDAon}
	\resizebox{.9\hsize}{!}{$
		\mathcal{F}_{c} = \frac{ 1 }{ G } \left[ 
			\sqrt{2} \csch^2(H)\Upsilon 
			+ 4 \coth(H) \frac{ 
			 4 \Sigma \cosh(2 H) + \cosh(3 H) \csch(H) + 4 H - 2 \Sigma
			 }{ 
				\mathcal{Q}_{H}
			} \Upsilon^2
			+ \mathcal{O}( \Upsilon^3 ) 
		\right]
	$}
\end{equation}
where $\mathcal{Q}_{H} = 2 \Sigma \cosh(2 H) + \sinh(2 H) + 2 H - 2 \Sigma$.

In the infinite-depth limit ($H\to\infty$), the first-order term vanishes, and the second order term collapses to $\frac{ 8 }{ G } \Upsilon^2$, in agreement with \eqref{eqn:CVAon}.

\subsection{The surfactant-covered infinite-depth case}
\label{ss:SurfLim}

We obtain the surfactant-covered infinite-depth case by letting $H \rightarrow \infty$ while keeping $\mathcal{M}^{\dagger}$ and $\mathcal{B}^{\dagger}$ arbitrary.
The onset acceleration becomes:
\begin{equation}
	\label{eqn:AonSurfInfiniteH}
	\begin{aligned} 
		\mathcal{F}_{c} = \frac{ 1 }{ G } \left[ 
			\sqrt{2} \left( \frac{ 
				\mathcal{Q}_S
				- 1 
				+ \frac{ \sqrt{2} \mathcal{M}^{\dagger} }{ 1 + \tfrac{1}{\mathcal{P}e^2} } 
			}{
				\mathcal{Q}_S
			} \right) \Upsilon 
			+ \left( 
				\frac{ 
					2 \Sigma \mathcal{N}_1
					+ \mathcal{N}_2
				}{ 
					\left( 2 \Sigma + 1 \right) \left( 1 + \tfrac{1}{\mathcal{P}e^2} \right)^3 \left( 
					\mathcal{Q}_S
					\right)^3
				} 
			\right) \Upsilon^2 
			+ \mathcal{O}( \Upsilon^3 ) 
		\right]
	\end{aligned} 
\end{equation}
where 
$$\mathcal{Q}_S = 1 + \sqrt{2} \mathcal{B}^{\dagger} + \mathcal{B}^{\dagger 2} + \frac{ \mathcal{M}^{\dagger} }{ \mathcal{P}e \left( 1 + \tfrac{1}{\mathcal{P}e^2} \right) } \left( \sqrt{2} + 2 \mathcal{B}^{\dagger} - \sqrt{2} \mathcal{P}e + \mathcal{M}^{\dagger} \mathcal{P}e \right) $$
and $\mathcal{N}_1$ and $\mathcal{N}_2$ (which are polynomials of $\mathcal{M}^{\dagger}$, $\mathcal{B}^{\dagger}$, $\tfrac{1}{\mathcal{P}e}$, and $1 + \tfrac{1}{\mathcal{P}e^2}$) are printed in appendix \ref{ass:InfDepthSurf}.

 	In the surfactant-free limit ($\mathcal{M}^{\dagger}\to0$ and $\mathcal{B}^{\dagger}\to0$), we find 
\begin{align*}
	\mathcal{Q}_S &\sim 1 \\
	\tfrac{\mathcal{N}_1}{ \left( 1 + \tfrac{1}{\mathcal{P}e^2} \right)^3 } \sim
	\tfrac{\mathcal{N}_2}{ \left( 1 + \tfrac{1}{\mathcal{P}e^2} \right)^3 } &\sim 8
\end{align*}
Thus the first-order term vanishes and the coefficient of the second-order term approaches $\frac{ 16 \Sigma + 8 }{ G ( 2 \Sigma + 1 ) } \Upsilon^2$.
Hence, equation \eqref{eqn:AonSurfInfiniteH} agrees with equation \eqref{eqn:CVAon} in this limit.

\subsection{The general solution}
\label{ss:GenSoln}

The general expression for the onset acceleration is:
\begin{equation}
	\label{eqn:AonGeneral}
	\resizebox{.9\hsize}{!}{$
	\begin{aligned} 
		\mathcal{F}_{c} =& \frac{ 1 }{ G } \left[ 
			\sqrt{2} \left( \frac{ 
				\mathcal{Q}_S \csch^2(H) 
				+ \left( \mathcal{Q}_S - 1 + \frac{ \sqrt{2} \mathcal{M}^{\dagger} }{ 1 + \tfrac{1}{\mathcal{P}e^2} } \right) \coth^2(H)
			}{
				\mathcal{Q}_S
			} \right) \Upsilon \right. \\
			& \hspace{2em} + \left( 
				\coth(H) \frac{ 
						\cosh(2 H) \left( 4 \Sigma \mathcal{L}_1 + \mathcal{L}_2 \coth(H) \right)
						+ \cosh(3 H) \csch(H) \mathcal{L}_3
						+ 4 H \mathcal{L}_4
						- 2 \Sigma \mathcal{L}_5
						+ \coth(H) \mathcal{L}_6
				}{ 
					\left( 1 + \tfrac{1}{\mathcal{P}e^2} \right)^3 \left( \mathcal{Q}_S \right)^3 \mathcal{Q}_H 
				} 
			\right) \Upsilon^2  \\
			& \left. \hspace{2em} + \mathcal{O}( \Upsilon^3 ) 
		\right]
	\end{aligned} 
	$}
\end{equation}
where the coefficients $\mathcal{L}_1$, $\mathcal{L}_2$, $\mathcal{L}_3$, $\mathcal{L}_4$, $\mathcal{L}_5$, and $\mathcal{L}_6$ (which are polynomials of $\mathcal{M}^{\dagger}$, $\mathcal{B}^{\dagger}$, $\tfrac{1}{\mathcal{P}e}$, and $1 + \tfrac{1}{\mathcal{P}e^2}$) are printed in appendix \ref{ass:GenSoln}.

In the surfactant-free limit, as $\mathcal{Q}_S \rightarrow 1$, the first order term approaches $\tfrac{\sqrt{2}\csch^2(H)}{G}\Upsilon$.
Noting that the $\mathcal{L}$ coefficients approach:
\begin{align*}
	\tfrac{\mathcal{L}_1}{\left( 1 + \tfrac{1}{\mathcal{P}e^2} \right)^3} \sim
	\tfrac{\mathcal{L}_3}{\left( 1 + \tfrac{1}{\mathcal{P}e^2} \right)^3} \sim
	\tfrac{\mathcal{L}_4}{\left( 1 + \tfrac{1}{\mathcal{P}e^2} \right)^3} \sim
	\tfrac{\mathcal{L}_5}{\left( 1 + \tfrac{1}{\mathcal{P}e^2} \right)^3} &\sim 4 \\
	\tfrac{\mathcal{L}_2}{\left( 1 + \tfrac{1}{\mathcal{P}e^2} \right)^3} \sim
	\tfrac{\mathcal{L}_6}{\left( 1 + \tfrac{1}{\mathcal{P}e^2} \right)^3} &\sim 0 
	,
\end{align*}
the second order term approaches $\tfrac{4\coth(H)}{G} \tfrac{4 \Sigma \cosh(2H) + \cosh(3H) \csch(H) + 4 H - 2 \Sigma}{\mathcal{Q}_H} \Upsilon^2$.
In this limit, the general solution agrees with eqn \ref{eqn:FDAon}.

	In the infinite-depth limit, the first order term approaches $\tfrac{\sqrt{2}}{G \mathcal{Q}_S} \left(\mathcal{Q}_S - 1 + \tfrac{\sqrt{2}\mathcal{M}^{\dagger}}{1 + \tfrac{1}{\mathcal{P}e^2}}\right)\Upsilon$.
Noting that $\mathcal{L}_1 = \tfrac{1}{2} \mathcal{N}_1$ and $\mathcal{L}_2 = \mathcal{N}_2 - 2 \mathcal{L}_3$, the second-order term asymptotically approaches 
$$\frac{ \tfrac{1}{2} e^{2 H} ( 2 \Sigma \mathcal{N}_1 + \mathcal{N}_2 - 2 \mathcal{L}_3 ) + e^{2 H} \mathcal{L}_3 }{ (1 + \tfrac{1}{\mathcal{P}e})^2 \mathcal{Q}_S^3 \tfrac{1}{2} ( 2 \Sigma + 1 ) e^{2 H} } ,$$
showing the general solution to agree with eqn \ref{eqn:AonSurfInfiniteH}.

\section{Comparing to previous numerical results}
\label{s:GiavedoniTest}

\begin{table}
	\centering
	\begin{tabular}{ c | c | c }
		Symbol & \citet{Kumar2004} & \citet{Giavedoni2007} \\
		\hline \hline
		$ \omega_0 $ (rad/s) & $\pi \times 60$ & $\pi \times 120$ \\
		$ \rho $ (kg/m$^3$) & $1 \times 10^{3}$ & $1 \times 10^{3}$ \\
		$ \mu $ (kg/m/s) & $1 \times 10^{-3}$ & $1 \times 10^{-3}$ \\
		$ \sigma_0 $ (kg/s$^2$) & $30 \times 10^{-2}$ & $70 \times 10^{-2}$ \\
		$ g $ (m/s$^2$) & $9.81$ & $9.80$ \\
		$ H $ (m) & $1 \times 10^{-2}$ & $1.5 \times 10^{-3}$ \\
		$ \varepsilon_0 $ (kg/s$^2$) & $0 - 10^{-1}$ & $10^{-6} - 10^{-1}$ \\
		$ D $ (m/s$^2$) & $10^{-9} - 10^{-2} $ & $10^{-9} - 10^{-3}$ \\
		$ \Omega $ (kg/s) & N/A & $0 - 10^{-3}$ \\
	\end{tabular}
	\caption{
		The parameter values and ranges used to compare our general solution to the numerical results of \citet{Kumar2004} and \citet{Giavedoni2007}.
		The results of this comparison are shown in figures \ref{f:KCompM} and \ref{f:GCompM}.
	}
	\label{t:params}
\end{table}

\begin{figure}
\centering
\includegraphics[width=0.85\linewidth]{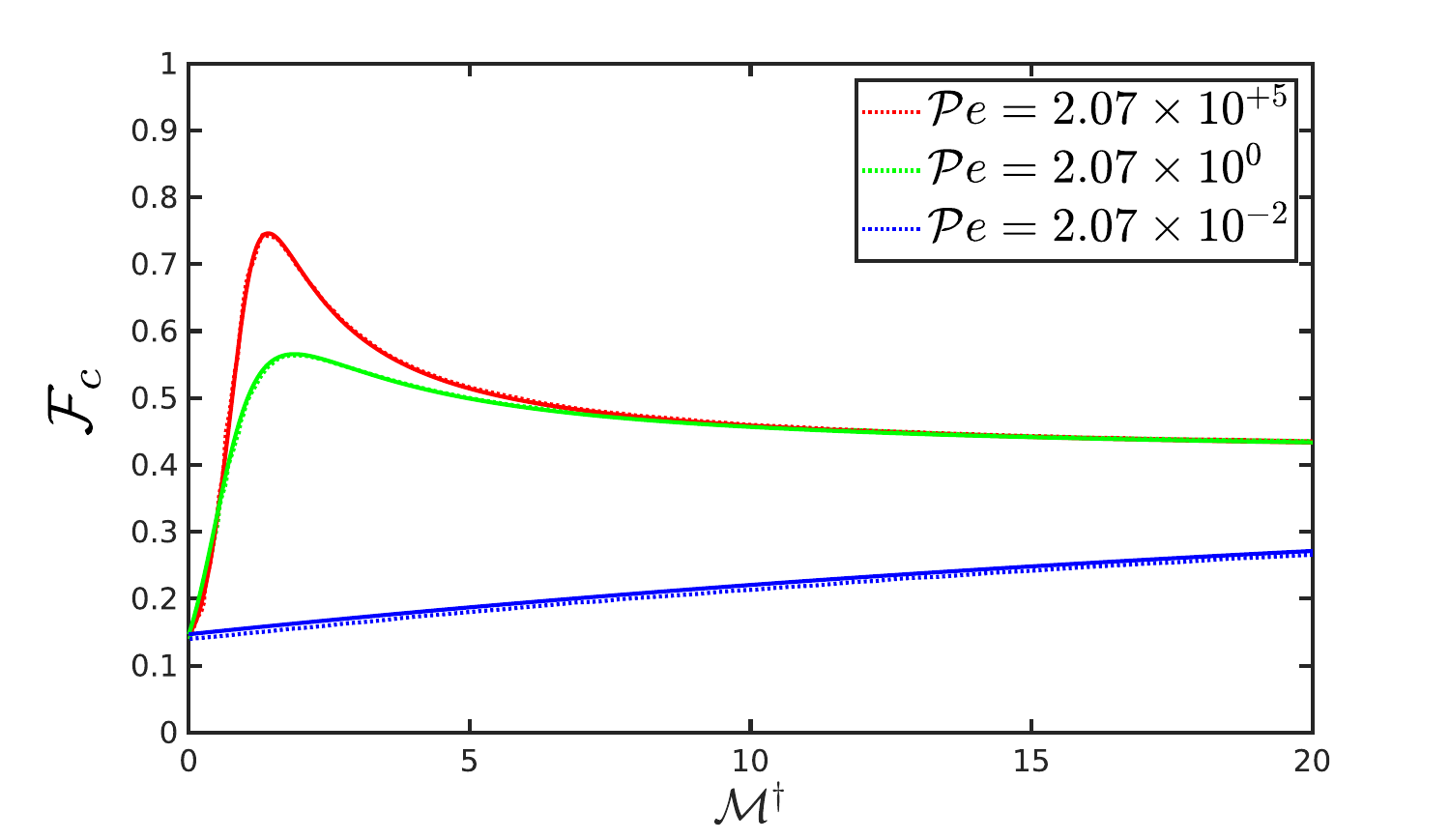}
\caption{
	{\bf Comparison of the analytic solution \eqref{eqn:AonGeneral} to numerical solutions of \citet{Kumar2004}}.
	The onset acceleration $\mathcal{F}_{c}$ is plotted against the modified Marangoni number $\mathcal{M}^{\dagger}$ for $\mathcal{B}^{\dagger}=0$ for a range of Peclet numbers $\mathcal{P}e$.
	The numerical results (finely dotted lines) are nearly indistinguishable from the corresponding analytic solution (solid lines).
}
\label{f:KCompM}
\end{figure}

\begin{figure}
\centering
\includegraphics[width=\linewidth]{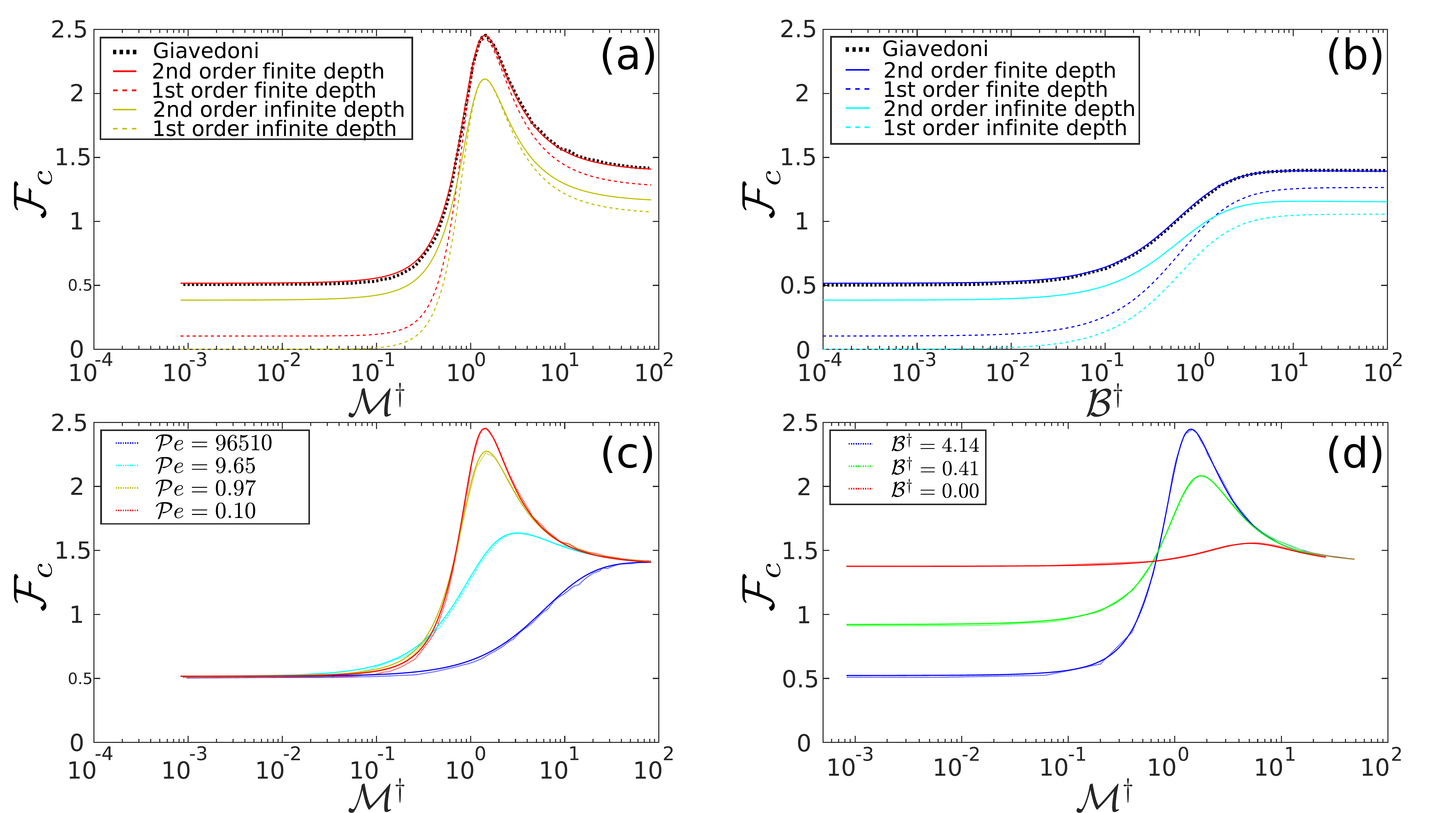}
\caption{
	{\bf Comparison of the analytic solution \eqref{eqn:AonGeneral} to numerical solutions of \citet{Giavedoni2007}}.
	{\bf (a)} $\mathcal{F}_{c}$ vs $\mathcal{M}^{\dagger}$ with $\mathcal{B}^{\dagger}=0$ and $\mathcal{P}e=7.991\times10^{4}$.
	The numerical result (finely dotted black line) is very close to the analytic solution \eqref{eqn:AonGeneral} (solid red line).
	The first-order term is shown as a dashed red curve.
	The corresponding infinite-depth analytic approximation \eqref{eqn:AonSurfInfiniteH} is shown as a solid yellow curve, and the first order contribution is a dashed yellow curve. 
	{\bf (b)} $\mathcal{F}_{c}$ vs $\mathcal{B}^{\dagger}$ with $\mathcal{M}^{\dagger}=0$ and $\mathcal{P}e=7.991\times10^{4}$.
	The numerical result (finely dotted black line) is very close to the analytic solution \eqref{eqn:AonGeneral} (solid blue line).
	The first-order term is shown as a dashed blue curve.
	The corresponding infinite-depth analytic approximation \eqref{eqn:AonSurfInfiniteH} is shown as a solid cyan curve, and the first order contribution is a dashed cyan curve.
%
	{\bf (c)} $\mathcal{F}_{c}$ vs $\mathcal{M}^{\dagger}$ for $\mathcal{B}^{\dagger}=0$ and a range of $\mathcal{P}e.$ 
	Numerical results are finely dotted curves, barely distiguishable from the corresponding analytic solutions \eqref{eqn:AonGeneral} shown as solid curves.
	{\bf (d)} $\mathcal{F}_{c}$ vs $\mathcal{M}^{\dagger}$ for $\mathcal{P}e=7.991\times10^{4}$ and a range of $\mathcal{B}^{\dagger}.$ 
	Numerical results are finely dotted curves, barely distiguishable from the corresponding analytic solutions \eqref{eqn:AonGeneral} shown as solid curves.
}
\label{f:GCompM}
\end{figure}

\begin{figure}
\centering
	\includegraphics[height=0.85\textheight]{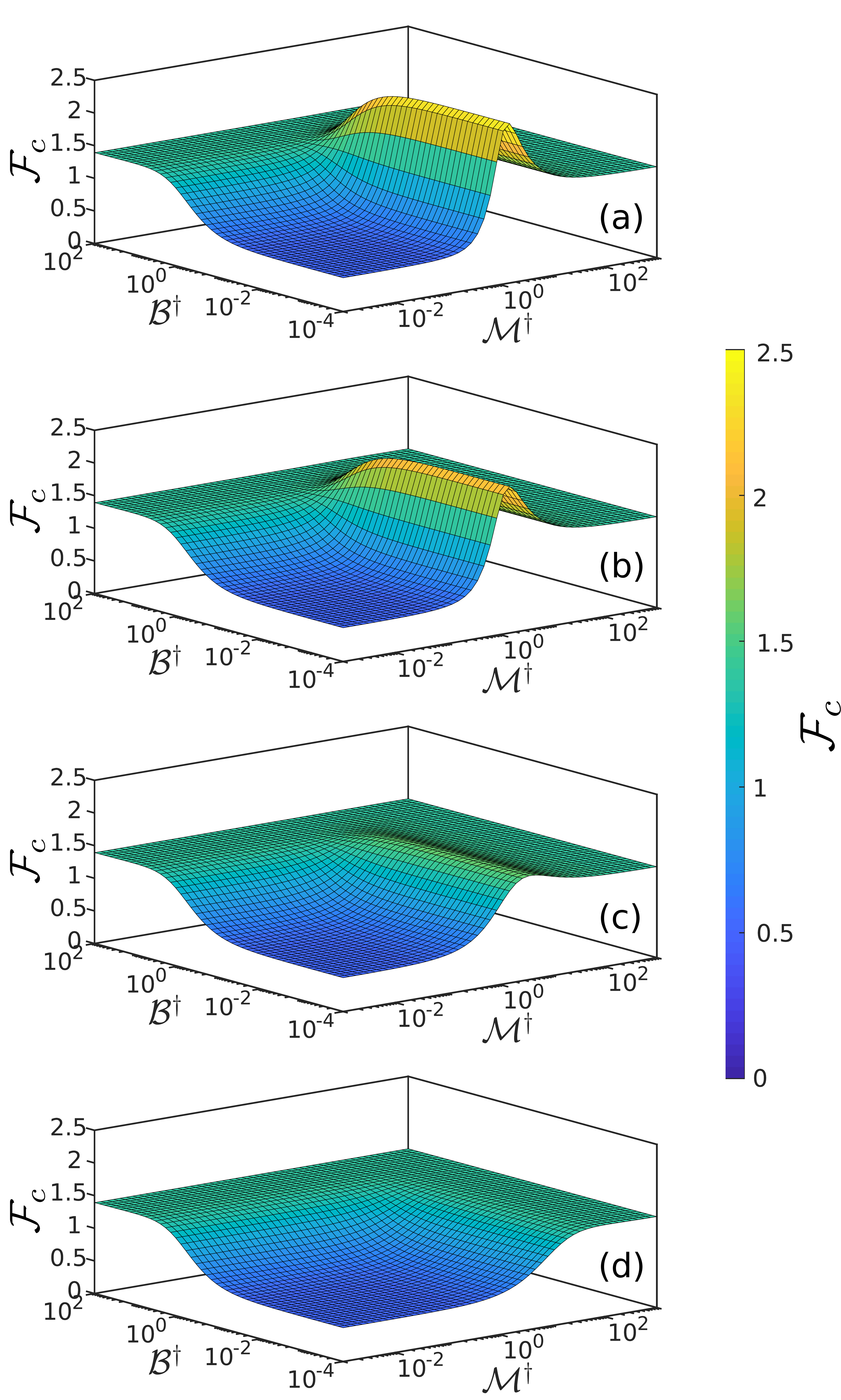}
\caption{
	{\bf Behavior of $\mathcal{F}_{c}$ in the $\mathcal{M}^{\dagger}$-$\mathcal{B}^{\dagger}$ plane} as the Peclet number is decreased: 
	(a) $\mathcal{P}e=7.991\times10^{4}$
	(b) $\mathcal{P}e=7.991\times10^{0}$
	(c) $\mathcal{P}e=7.991\times10^{-1}$
	(d) $\mathcal{P}e=7.991\times10^{-2}$.
	These graphs show the role that surface diffusion plays in reducing, broadening, and moving the maximum to higher $\mathcal{M}^{\dagger}$.
	All of these graphs consider a finite-depth water-like bulk fluid using the same physical parameters as Fig.~\ref{f:GCompM}.
}
\label{f:avsMBP}
\end{figure}

\citet{Kumar2004,Kumar2004a,Ubal2005,Ubal2005b}, and \citet{Giavedoni2007} studied the finite-depth surfactant-covered Faraday wave problem, examining the effect of surfactants on the onset acceleration, wave number, and the phase shift between the surfactant distribution and the surface topography.
\citet{Kumar2004,Kumar2004a} and \citet{Ubal2005}  accounted for Marangoni and Peclet effects while \citet{Ubal2005b} considered Boussinesq effects only.
\citet{Giavedoni2007} generalize results of  \citet{Ubal2005} and \citet{Ubal2005b}, accounting for all three effects.
In establishing the efficacy of our analytic solution, we compare to \citet{Kumar2004,Kumar2004a} and \citet{Giavedoni2007}.
The parameters for these studies are shown in Table \ref{t:params}, and include a wide range of surface elasticities, viscosities, and diffusivities but only a few values for the fluid depth, viscosity, surface tension, density, and driving frequency.

Before comparing results, it is worth contrasting our analytic approach to the recursion relation in eqn \ref{eqn:RecursRelat} with previous numerical approaches.
In our analysis, we truncated the recursion relation at an arbitrarily large $n$, established a base case, and Taylor-expanded the problem in the weak-viscosity limit to second order in $\Upsilon$; the numerical approaches are based on truncating the relation at  $j=10$, casting the relation into matrix form, and numerically solving the remaining expression as an eigenvalue problem, deducing the onset acceleration from the eigenvalue.
Neither \citet{Kumar2004} nor \citet{Giavedoni2007} explicitly considered the weak-viscosity limit, but based on the parameters reported, they studied $\Upsilon = 0.0696$ and $\Upsilon = 0.0644$ respectively, well within the weak-viscosity limit.

In order to compare results, we converted the $\mathcal{M}$ and $\mathcal{B}$ from these numerical studies first to physical parameter values $\varepsilon_0$, $D$, and $\Omega$ and then into the $\mathcal{M}^{\dagger}$ and $\mathcal{B}^{\dagger}$ used here.
This process requires careful accounting, not only in the additional use of $\Upsilon$ in the definitions of $\mathcal{M}^{\dagger}$ and $\mathcal{B}^{\dagger}$, but also in the definitions of the length and time scales used to non-dimensionalize the problem.
Although we use the wavelength from the finite-depth Kelvin dispersion relation as the length scale, \citet{Kumar2004} used the fluid depth $H$, and \citet{Giavedoni2007} used $l_0 = \tfrac{g}{\omega^2} + \sqrt[3]{\tfrac{\sigma_0}{\rho \omega^2}}$.

Figure \ref{f:KCompM} shows that our analysis is in quantitative agreement with the results of \citet{Kumar2004} for a wide range of $\mathcal{M}^{\dagger}$ and $\mathcal{P}e$.
These curves clearly show that when diffusion is negligible, the onset acceleration rapidly increases with surface elasticity up to a maximum that is significantly larger than the surfactant-free case.
The onset acceleration then decreases to nearly half its peak value, very similar to the energy damping rate in linear gravity-capillary waves as discussed in section \S\ref{s:Intro}.
Surface diffusivity acts to reduce, broaden, and shift the peak to higher values of Marangoni number.

Figure \ref{f:GCompM} shows that our analysis also agrees with the results of \citet{Giavedoni2007} for five decades of $\mathcal{M}^{\dagger}$, eight decades of $\mathcal{B}^{\dagger}$, and six decades of $\mathcal{P}e$.
Figure \ref{f:GCompM} (a) and (c) show the dependence of $\mathcal{F}_{c}$ vs $\mathcal{M}^{\dagger}$ and $\mathcal{P}e$, and although the fluid depth, driving frequency, and surface tension are different than in Fig. \ref{f:KCompM}, the trends in the plots are the same.
Figure \ref{f:GCompM} (b) shows $\mathcal{F}_{c}$ vs $\mathcal{B}^{\dagger}$ 
which exhibits a steady rise to a plateau.
Figure \ref{f:GCompM} (d) shows how surface viscosity also reduces, broadens, and moves the peak in the $\mathcal{F}_{c}$ vs $\mathcal{M}^{\dagger}$ plot.
The surface viscosity also increases the onset acceleration at low $\mathcal{M}^{\dagger}$.

To help visualize the behavior of $\mathcal{F}_{c}$ in the $\mathcal{M}^{\dagger}$-$\mathcal{B}^{\dagger}$ plane, we have added figure \ref{f:avsMBP}.
Although no new comparison is made in this figure, it uses the same physical parameters as Fig.~\ref{f:GCompM}.
This visualization clearly shows that  $\lim_{\mathcal{M}^{\dagger} \rightarrow \infty} \mathcal{F}_{c} = \lim_{\mathcal{B}^{\dagger} \rightarrow \infty} \mathcal{F}_{c}$.
Further, as $\mathcal{P}e \rightarrow 0$, the peak lessens, broadens, shifts to higher values of $\mathcal{M}^{\dagger}$, and ultimately vanishes.

Figure \ref{f:GCompM} (a,b) additionally show the behavior of the first-order and second-order terms of the analytic solution \eqref{eqn:AonGeneral} and infinite-depth analytic approximation \eqref{eqn:AonSurfInfiniteH}.
The first-order terms captures the overall trends, particularly the maximum when $\mathcal{M}^{\dagger} \approx 1$ and the rise when $\mathcal{B}^{\dagger} \approx 0.5$, but generally under-predict the onset acceleration.
The second-order terms correct the under-prediction.

\section{Additional features of the analytic solution}
\label{s:ExploreAnalyticalSoln}

\begin{figure}
\centering
	\includegraphics[width=0.85\textwidth]{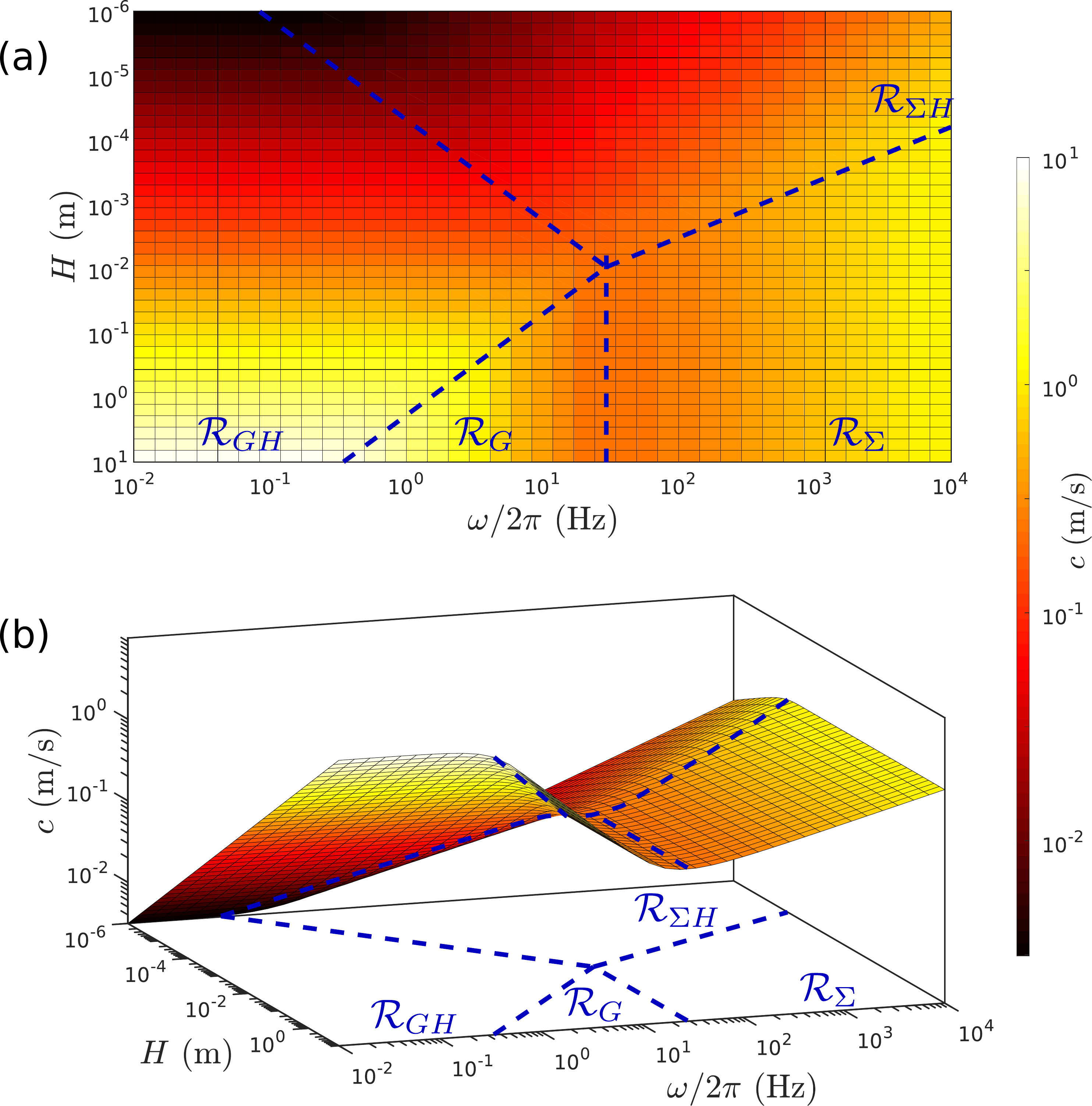}
\caption{
	{\bf Finite-depth Kelvin dispersion relation wave speed $c = l_0 \omega_0$ in the $H$-$\omega$ plane} using water-like conditions where the values for the physical parameters are: $\rho=1\times10^{3}$~$\tfrac{\text{kg}}{\text{m}^3}$, $\mu = 1\times10^{-3}$~$\tfrac{\text{kg}}{\text{m s}}$, $\sigma_0=70\times10^{-3}$~$\tfrac{\text{N}}{\text{m}}$, and $g=9.8$~$\tfrac{\text{m}}{\text{s}^2}$.
	These plots show the same dispersion relation from two different perspectives.
	Fig (a) is a top view of the $H$-$\omega$ plane while fig (b) uses the same perspective of the plane as shown in figures \ref{f:avsHOmegaRegions},\ref{f:avsHOmega-Elasticity},\ref{f:avsHOmega-Viscosity}.
	A few regions have been indicated:
	$\mathcal{R}_{G}$ are gravity waves, 
	$\mathcal{R}_{G H}$ are depth-restricted gravity waves, 
	$\mathcal{R}_{\Sigma}$ are capillary waves, 
	and $\mathcal{R}_{\Sigma H}$ are depth-restricted capillary waves.
}
\label{f:KelvinDispersionRelation}
\end{figure}

\begin{figure}
\centering
	\includegraphics[width=0.85\textwidth]{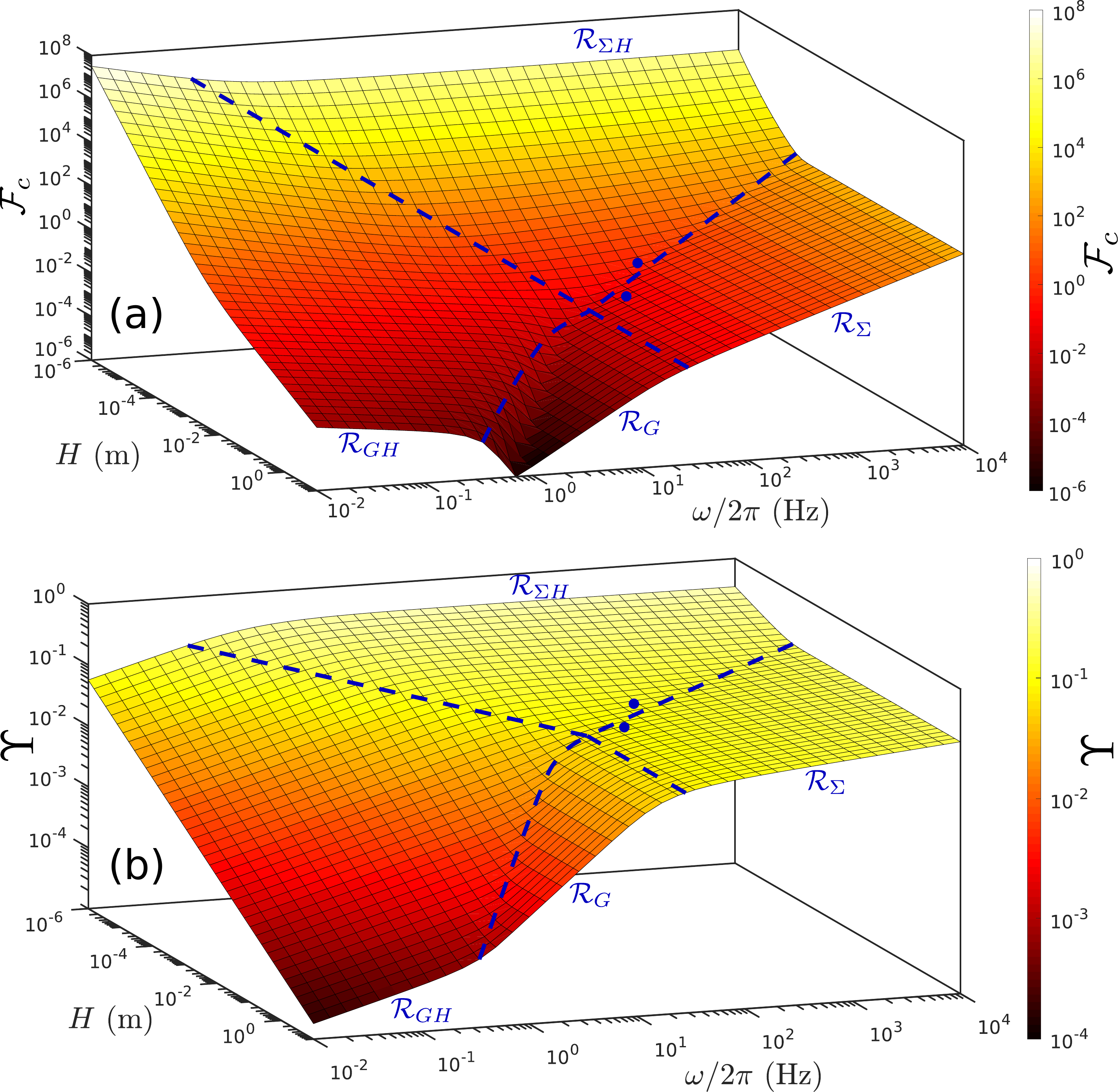}
\caption{
	{\bf Behavior of $\mathcal{F}_{c}$ and the expansion parameter $\Upsilon$ in the $H$-$\omega$ plane} using the same surfactant-free water-like conditions as in figure \ref{f:KelvinDispersionRelation}.
	The same regions from fig \ref{f:KelvinDispersionRelation} are evident in the behavior of $\mathcal{F}_{c}$ in that at the boundaries of each region, $\mathcal{F}_{c}$ exhibits significant curvature.
	The two blue dots indicate where our comparison with \citet{Kumar2004} and \citet{Giavedoni2007} occurred.
}
\label{f:avsHOmegaRegions}
\end{figure}

\begin{figure}
\centering
	\includegraphics[width=\textwidth]{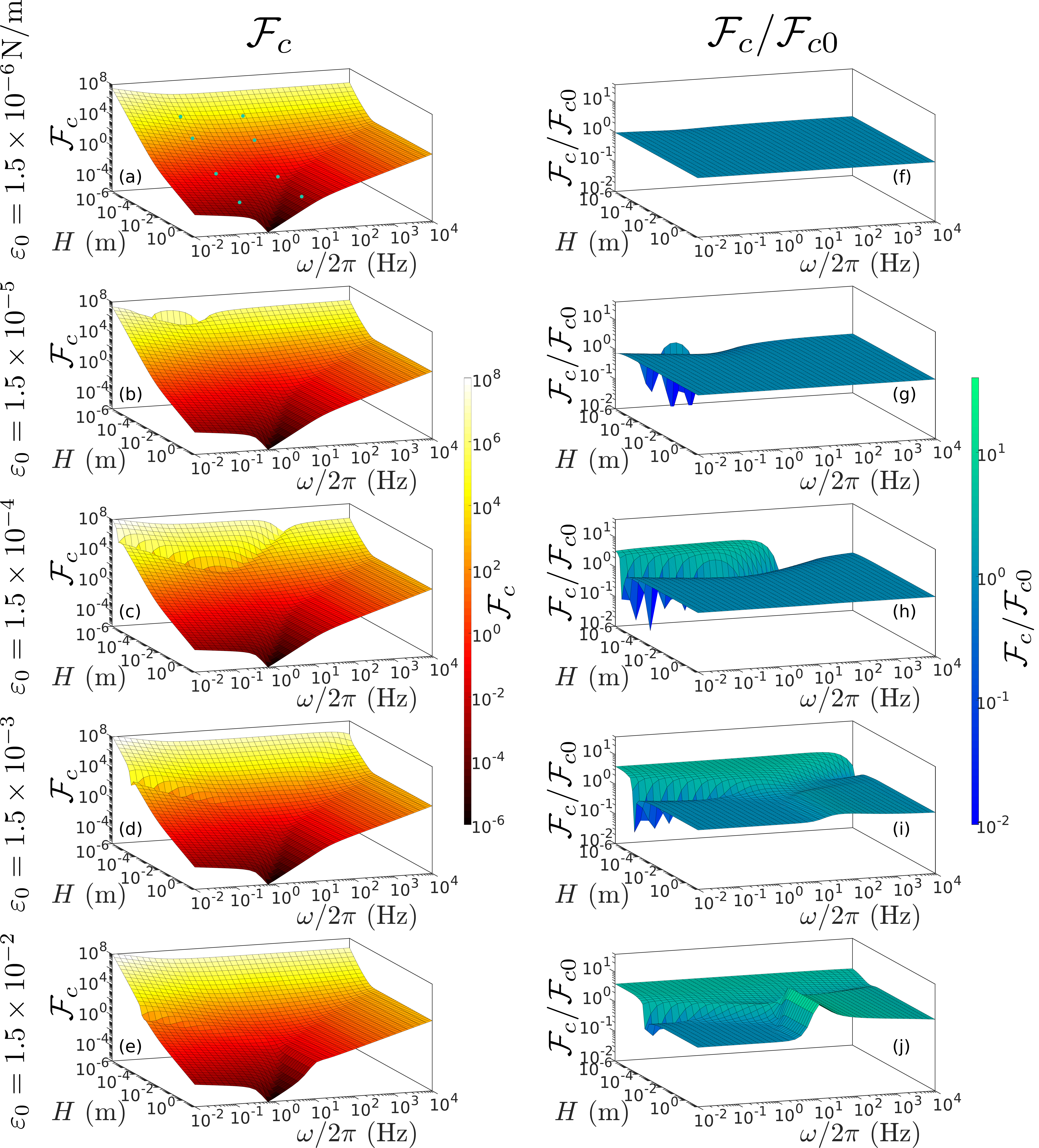}
\caption{
	{\bf Behavior of $\mathcal{F}_{c}$ in the $H$-$\omega$ plane as surface elasticity is increased}.
	These figures consider the same water-like conditions as in figure \ref{f:KelvinDispersionRelation} and \ref{f:avsHOmegaRegions} but now with a surfactant that only affects the surface elasticity and does not diffuse.
	The left column of figures show $\mathcal{F}_{c}$ while the right column shows the ratio of $\mathcal{F}_{c}$ to $\mathcal{F}_{c0}$ (the corresponding surfactant-free onset acceleration from fig.~\ref{f:avsHOmegaRegions}).
	A ratio of $10^{0} = 1$ means that the onset acceleration is indistinguishable from the surfactant-free case.
	The eight cyan dots indicate locations on the $H$-$\omega$ plane corresponding to the subplots in figure \ref{f:avsHOmega-MBPlanes}.
}
\label{f:avsHOmega-Elasticity}
\end{figure}

\begin{figure}
\centering
	\includegraphics[width=\textwidth]{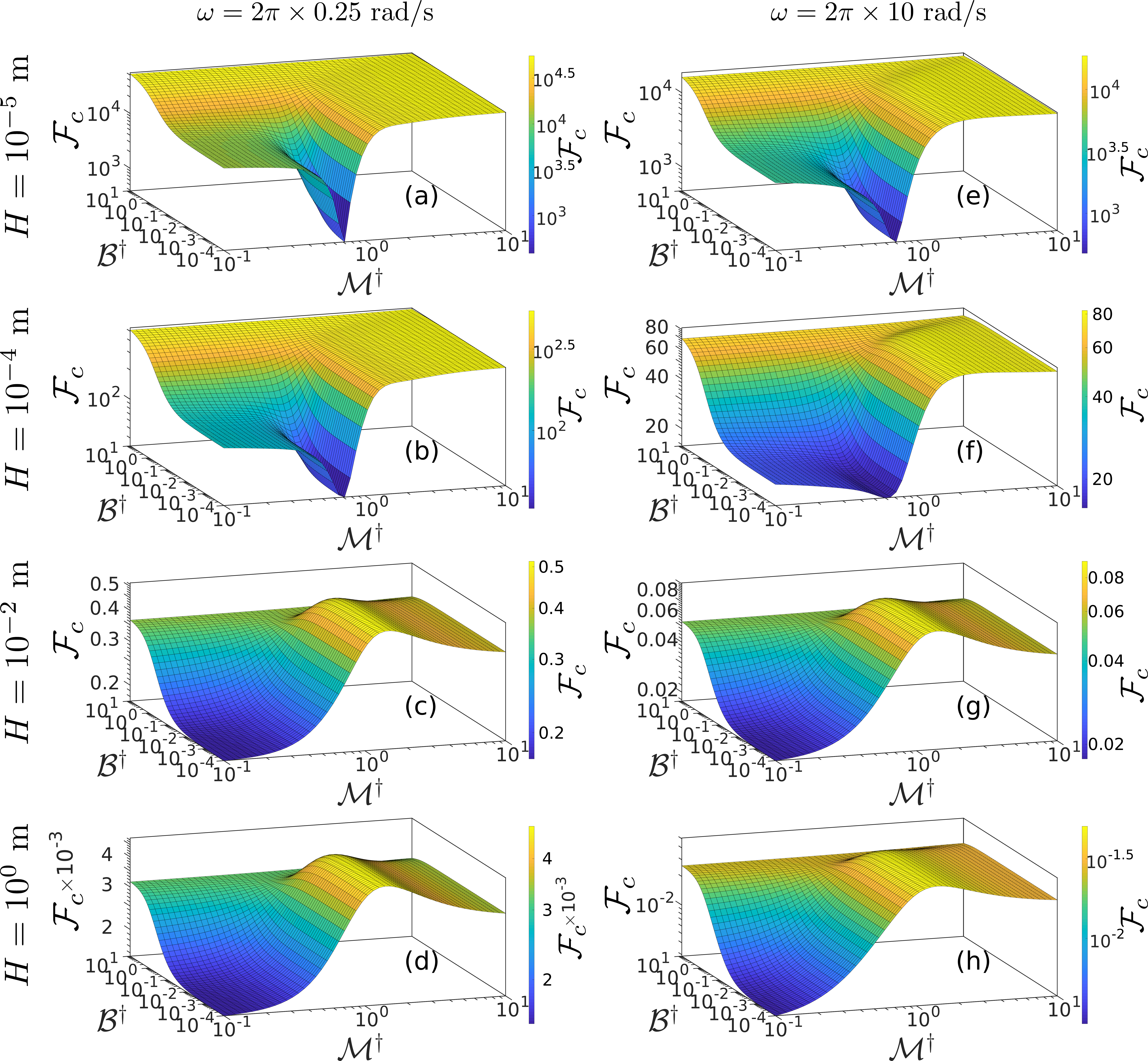}
\caption{
	{\bf Behavior of $\mathcal{F}_{c}$ in the $\mathcal{M}^{\dagger}$-$\mathcal{B}^{\dagger}$ plane} for several $H$ and $\omega$.
	Each plot corresponds to a cyan dot in fig \ref{f:avsHOmega-Elasticity} (a).
	Each column corresponds to a driving frequency and each row corresponds to a fluid depth.
	The left column has a frequency of $\omega = 2 \pi \times 0.25$~rad/s while the right has $\omega = 2 \pi \times 10$~rad/s.
	The depths of each row is:
	(a,e) $H = 1 \times 10^{-5}$~m 
	(b,f) $H = 1 \times 10^{-4}$~m
	(c,g) $H = 1 \times 10^{-2}$~m
	(d,h) $H = 1 \times 10^{0}$~m.
	Although all of the plots use the same coloration, the color scaling is unique for each plot as indicated by the colorbars.
}
\label{f:avsHOmega-MBPlanes}
\end{figure}

\begin{figure}
\centering
	\includegraphics[width=\textwidth]{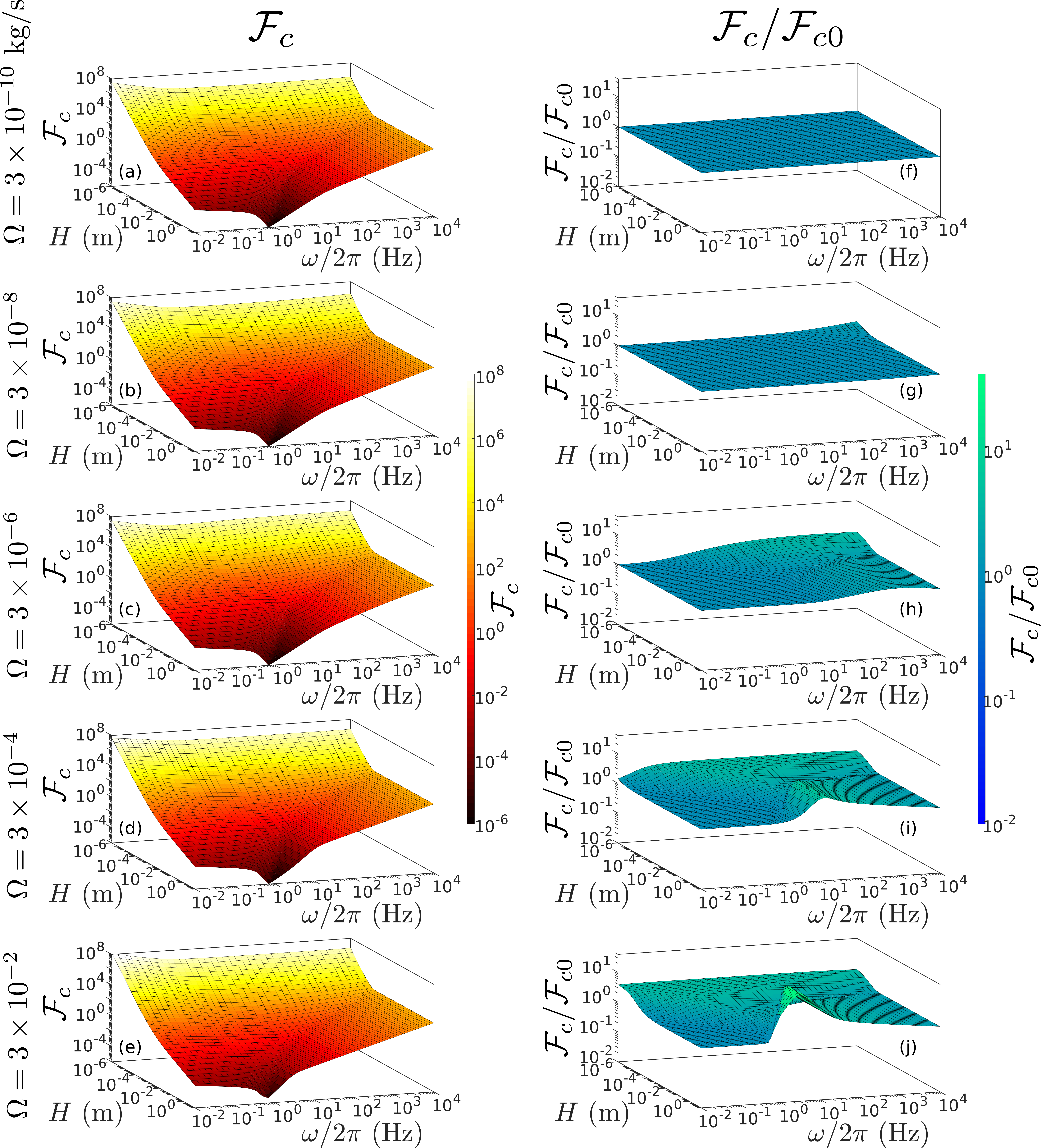}
\caption{
	{\bf Behavior of $\mathcal{F}_{c}$ in the $H$-$\omega$ plane as surface viscosity is increased}.
	These figures consider the same water-like conditions as in figures \ref{f:KelvinDispersionRelation}, \ref{f:avsHOmegaRegions}, and \ref{f:avsHOmega-Elasticity} but now with a surfactant that only affects the surface viscosity and does not diffuse.
	The left column of figures show $\mathcal{F}_{c}$ while the right column shows the ratio of $\mathcal{F}_{c}$ to $\mathcal{F}_{c0}$ (the corresponding surfactant-free onset acceleration from fig.~\ref{f:avsHOmegaRegions}).
	A ratio of $10^{0} = 1$ means that the onset acceleration is indistinguishable from the surfactant-free case.
}
\label{f:avsHOmega-Viscosity}
\end{figure}

The results presented in section \S\ref{s:GiavedoniTest} cover a wide range of $\mathcal{M}^{\dagger}$, $\mathcal{B}^{\dagger}$, and $\mathcal{P}e$.
Table \ref{t:params} shows that this range was achieved by varying the surface parameters, surface elasticity, surface viscosity, and surface diffusivity, while keeping constant the parameters for the bulk fluid.
Because of our analytic treatment, we are able to efficiently explore the behavior of $\mathcal{F}_{c}$ for a wide range of surface and bulk parameters.
In this section, we will consider the behavior of $\mathcal{F}_{c}$ in the $H$-$\omega$ plane, and observe that in some regions of this plane, increasing $\mathcal{M}^{\dagger}$ can significantly decrease the onset acceleration.
We will also explore the dependence of $\mathcal{F}_{c}$ on $\mu$, $\rho$, and $\sigma_0$.

The $H$ dependence of $\mathcal{F}_{c}$ (and similarly for the $\omega$ dependence) arises in two distinct ways.
Although the general solution (eqn \ref{eqn:AonGeneral}) explicitly references $H$, the fluid depth also affects the finite-depth Kelvin dispersion relation which is used to determine the characteristic length scale $l_0$ which is incorporated into nearly all of the dimensionless numbers.
Consequently, probing the depth-dependence cannot be done by merely plotting eqn \ref{eqn:AonGeneral} while holding the dimensionless numbers constant.
Similarly, $\omega$ directly contributes to the dimensionless numbers via $\omega_0$, and it also affects the dimensionless numbers through $l_0$ via the same dispersion relation.

To illuminate the role of the finite-depth Kelvin dispersion relation, fig \ref{f:KelvinDispersionRelation} shows the wave speed $c = l_0 \omega_0$ in the $H$-$\omega$ plane for terrestrial water ($\rho = 1000$~kg/m$^3$, $\mu = 0.001$~kg/m/s, $\sigma_0 = 0.07$~N/m, $g = 9.8$~m/s$^2$).
This dispersion relation strictly applies to linear gravity-capillary waves, but in the limit $\Upsilon \rightarrow 0$, the dispersion relation for non-linear Faraday waves approaches these plots.
These plots clearly show the gravity wave region $\mathcal{R}_{G}$ and the capillary wave region $\mathcal{R}_{\Sigma}$ as well as two new regions which we refer to as depth-restricted gravity waves $\mathcal{R}_{G H}$ and depth-restricted capillary waves $\mathcal{R}_{\Sigma H}$ since the finite depth of the container results in slower waves.
This dispersion relation is so significant that all of these regions are apparent in the behavior of $\mathcal{F}_{c}$ in the $H$-$\omega$ plane.

Figure \ref{f:avsHOmegaRegions} (a) shows $\mathcal{F}_{c}$ in the $H$-$\omega$ plane for surfactant-free water.
In later figures, we use this surfactant-free onset acceleration as a reference and denote it as $\mathcal{F}_{c0}$.
In this figure, we have marked the four regions from figure \ref{f:KelvinDispersionRelation} on the plot.
Although the most eye-catching feature is the valley that traces along the $\mathcal{R}_{G}$-$\mathcal{R}_{G H}$ border, each border either coincides with or is next to significant curvature.
Since the plot is logarithmic on all axes, any planar surfaces indicate power-law behavior, and any curvature from a plane corresponds to a change in the exponent.
Figure \ref{f:avsHOmegaRegions} (b) shows the low-viscosity expansion parameter $\Upsilon$ in the same $H$-$\omega$ plane.
In considering water, all points shown on the $H$-$\omega$ plane have $\Upsilon<1$ with the maximum value of $\Upsilon = 0.3458$ occurring at the shallowest depth and highest driving frequency.
The two blue dots in figure \ref{f:avsHOmegaRegions} show the locations in the $H$-$\omega$ plane where we compared our analysis to \citet{Kumar2004} and \citet{Giavedoni2007} which lie in the $\mathcal{R}_{\Sigma}$ and $\mathcal{R}_{\Sigma H}$ regions respectively.

Figure \ref{f:avsHOmega-Elasticity} shows the effect of surface elasticity on the onset acceleration by showing a progression of graphs of $\mathcal{F}_{c}$ and $\tfrac{\mathcal{F}_{c}}{\mathcal{F}_{c0}}$ in the $H$-$\omega$ plane.
For small elasticities, a new wedge-shaped region appears at shallow depths and mid-range driving frequencies.
We will refer to this region as the elasticity-affected region $\mathcal{R}_{\varepsilon}$.
As the surface elasticity increases, $\mathcal{R}_{\varepsilon}$ descends the graph, pressing towards deeper depths.
The tip of the wedge follows the $\mathcal{R}_{G H}$-$\mathcal{R}_{\Sigma H}$ boundary all the way to the quadruple point where the four types of waves meet.
Within $\mathcal{R}_{\varepsilon}$, the onset acceleration are elevated above the surfactant-free behavior, but at the boundary, the onset acceleration may either increase or decrease.
Figure \ref{f:avsHOmega-Elasticity} shows that the boundary tends to decrease the onset acceleration for shallow systems and increase the onset acceleration for deep systems.
Figure \ref{f:avsHOmega-MBPlanes} shows this unusual effect of surface elasticity in more detail by plotting $\mathcal{F}_{c}$ in the $\mathcal{M}^{\dagger}$-$\mathcal{B}^{\dagger}$ plane for eight different locations of the $H$-$\omega$ plane, the $H$-$\omega$ location of each plot is shown in figure \ref{f:avsHOmega-Elasticity} (a).
Each column corresponds to a driving frequency and each row corresponds to a fluid depth.
Notably, the application of a surfactant can decrease the onset acceleration by more than an order of magnitude.
Although the effects of diffusion are not shown in fig \ref{f:avsHOmega-Elasticity}, increasing the diffusivity only results in a lessening of the Marangoni-induced extremes (peaks and valleys) and a broadening of the boundary around $\mathcal{R}_{\varepsilon}$.
Diffusivity does not affect the dependence of the onset acceleration on the Boussinesq number.

Figure \ref{f:avsHOmega-Viscosity} shows the effect of surface viscosity on the onset acceleration by showing a progression of graphs of $\mathcal{F}_{c}$ and $\tfrac{\mathcal{F}_{c}}{\mathcal{F}_{c0}}$ in the $H$-$\omega$ plane.
Unlike surface elasticity, increasing surface viscosity will only increase the onset acceleration.
Weak surface viscosities will elevate the onset acceleration at high frequency across all depths, and further increasing the surface viscosity shifts these effects to lower frequencies.

\begin{figure}
\centering
	\includegraphics[width=\textwidth]{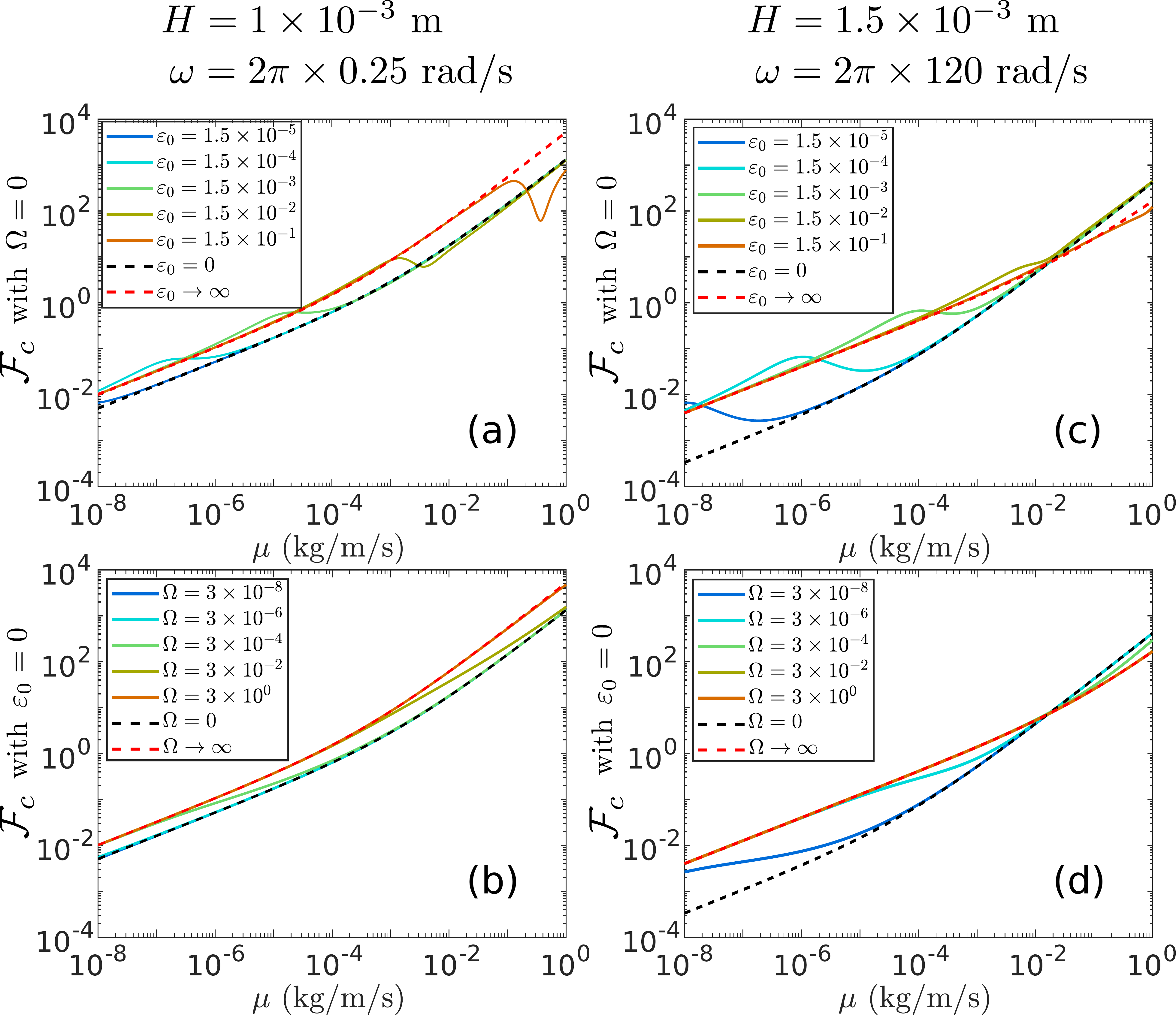}
\caption{
	{\bf Dependence of $\mathcal{F}_{c}$ on the bulk viscosity}.
	Similar to the parameters used in \citet{Giavedoni2007}, the system parameters are $\rho = 1 \times 10^3$~kg/m$^3$, $\sigma_0 = 70 \times 10^{-3}$~N/m, and $g = 9.8$~m/s$^2$.
	(a,b) consider depth-restricted gravity waves where $H = 1 \times 10^{-3}$~m and $\omega = 2 \pi \times 0.25$~rad/s while (c,d) consider depth-restricted capillary waves where $H = 1.5 \times 10^{-3}$~m and $\omega = 2 \pi \times 120$~rad/s.
	(a,c) consider a range of $\varepsilon_0$ with $\Omega = 0$ while (b,d) consider a range of $\Omega$ with $\varepsilon_0 = 0$.
	The limiting case of surfactant-free is shown by a dashed black line while the case of a surfactant-saturated surface is shown by a red dashed line.
}
	\label{f:avsMu}
\end{figure}

\begin{figure}
\centering
	\includegraphics[width=\textwidth]{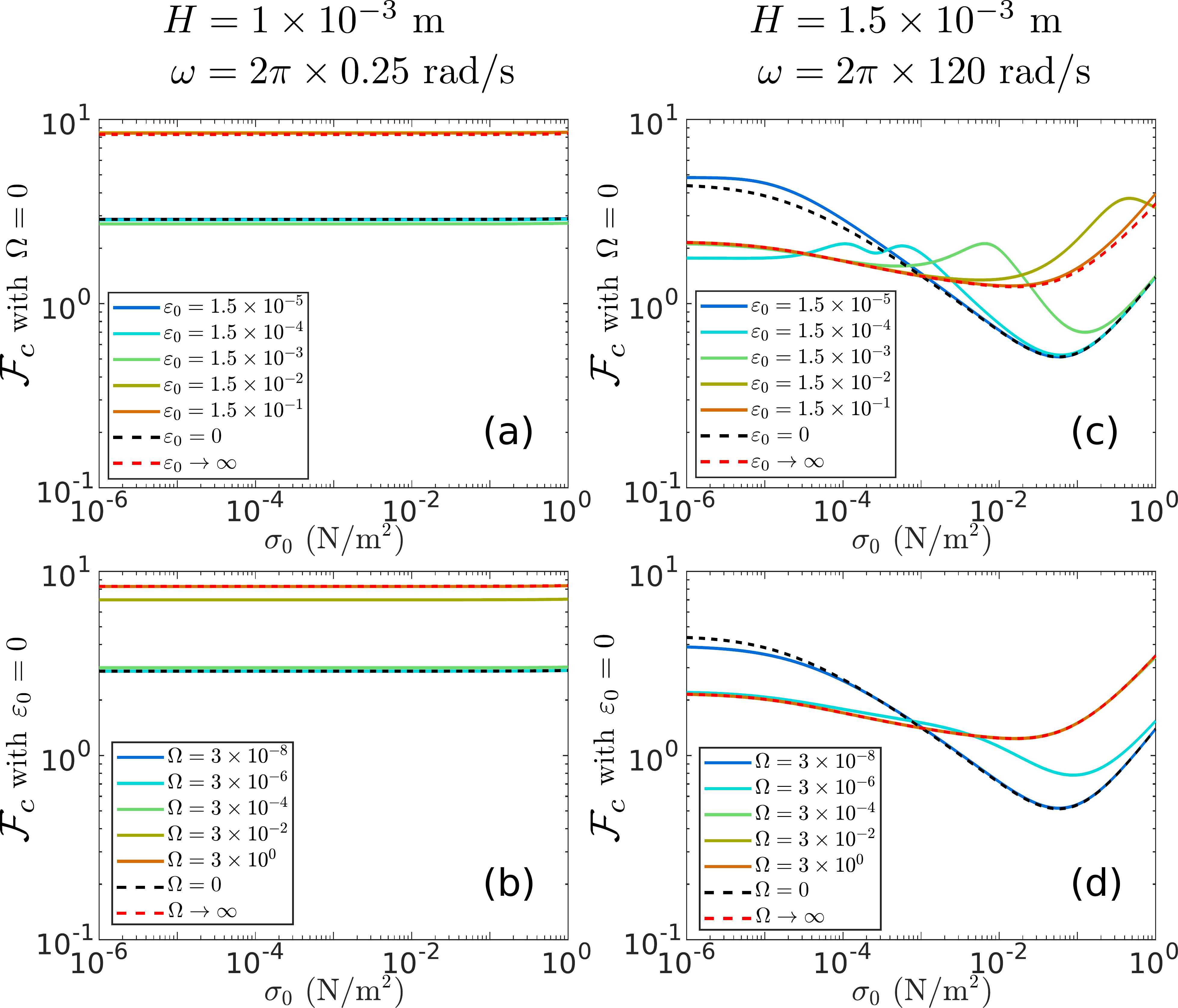}
\caption{
	{\bf Dependence of $\mathcal{F}_{c}$ on the surface tension}.
	Similar to the parameters used in \citet{Giavedoni2007}, the system parameters are $\rho = 1 \times 10^3$~kg/m$^3$, $\mu = 1 \times 10^{-3}$~kg/m/s, and $g = 9.8$~m/s$^2$.
	(a,b) consider depth-restricted gravity waves where $H = 1 \times 10^{-3}$~m and $\omega = 2 \pi \times 0.25$~rad/s while (c,d) consider depth-restricted capillary waves where $H = 1.5 \times 10^{-3}$~m and $\omega = 2 \pi \times 120$~rad/s.
	(a,c) consider a range of $\varepsilon_0$ with $\Omega = 0$ while (b,d) consider a range of $\Omega$ with $\varepsilon_0 = 0$.
	The limiting case of surfactant-free is shown by a dashed black line while the case of a surfactant-saturated surface is shown by a red dashed line.
}
	\label{f:avsSigma}
\end{figure}

\begin{figure}
\centering
	\includegraphics[width=\textwidth]{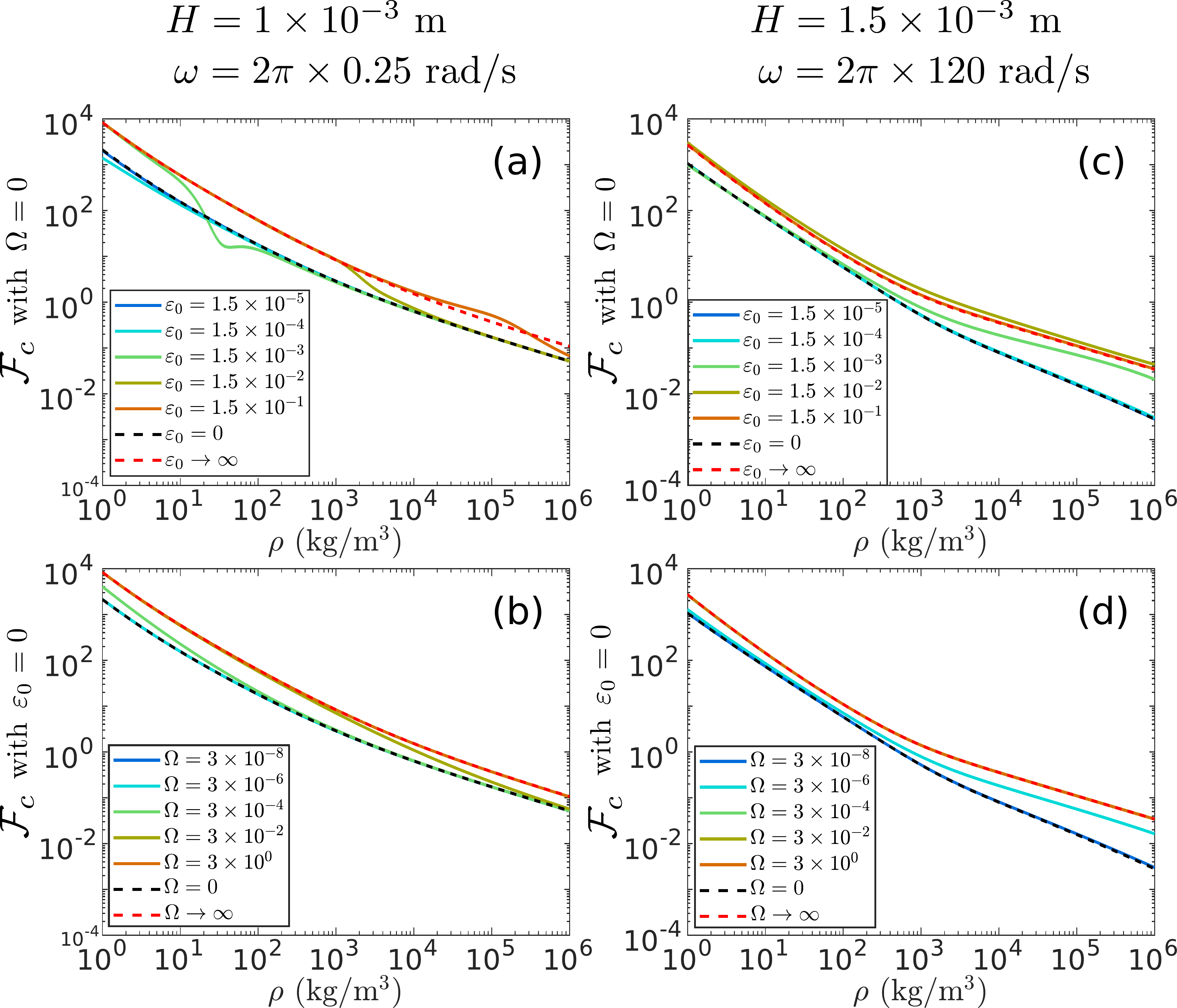}
\caption{
	{\bf Dependence of $\mathcal{F}_{c}$ on the bulk density}.
	Similar to the parameters used in \citet{Giavedoni2007}, the system parameters are $\mu = 1 \times 10^{-3}$~kg/m/s, $\sigma_0 = 70 \times 10^{-3}$~N/m, and $g = 9.8$~m/s$^2$.
	(a,b) consider depth-restricted gravity waves where $H = 1 \times 10^{-3}$~m and $\omega = 2 \pi \times 0.25$~rad/s while (c,d) consider depth-restricted capillary waves where $H = 1.5 \times 10^{-3}$~m and $\omega = 2 \pi \times 120$~rad/s.
	(a,c) consider a range of $\varepsilon_0$ with $\Omega = 0$ while (b,d) consider a range of $\Omega$ with $\varepsilon_0 = 0$.
	The limiting case of surfactant-free is shown by a dashed black line while the case of a surfactant-saturated surface is shown by a red dashed line.
}
	\label{f:avsRho}
\end{figure}

Figures \ref{f:avsMu}, \ref{f:avsSigma}, and \ref{f:avsRho} show the effects of bulk viscosity, surface tension, and bulk density respectively.
The parameters are based on water where $\mu = 10^{-3}$~kg/m/s, $\sigma_0 = 70 \times 10^{-3}$~N/m, and $\rho = 10^{3}$~kg/m$^3$, and in each figure, we vary a single parameter.
Frames (a,b) of these figures consider depth-restricted gravity waves ($\omega = 2 \pi \times 0.25$~rad/s and $H = 10^{-3}$~m), while frames (c,d) consider depth-restricted capillary waves ($\omega = 2 \pi \times 120$~rad/s and $H = 1.5 \times 10^{-3}$~m).

Figure \ref{f:avsMu} graphs the onset acceleration $\mathcal{F}_{c}$ vs the bulk viscosity $\mu$ showing that increasing the bulk viscosity generally increases the onset acceleration.
Although the Boussinesq effects shown in frames (b,d) always increase the onset acceleration, the Marangoni effects shown in frames (a,c) will sometimes decrease the onset acceleration as noted before.
Further, the presence of surface elasticity can result in cases where a more viscous fluid could have a lower onset acceleration than a less viscous fluid.
These figures also show the extreme cases of a surfactant-free fluid (black dashed line) and a surfactant-saturated fluid (red dashed line).
For gravity waves, a surfactant-saturated fluid always has a higher onset acceleration than a surfactant-free fluid; however, for capillary waves, there is a crossover where a saturated surface will onset Faraday waves more readily than a surfactant-free surface.
This crossover is for large viscosities, and since this model is designed for the low-viscosity limit and only tested against numerical results using a viscosity of $10^{-3}$~kg/m/s, the crossover may be a limitation of our second-order analysis.
However, \citet{Suman2008}  reported similar numerical results for high-viscosity systems, finding that the onset acceleration for an inertial-less surfactant-free system would be infinite, but Marangoni stresses allow Faraday waves to emerge, thereby preferentially decreasing the onset acceleration.

Fig \ref{f:avsSigma} graphs the onset acceleration $\mathcal{F}_{c}$ vs equilibrium surface tension $\sigma_0$.
The gravity-waves in frames (a,b) are unaffected by surface tension while the capillary waves in frames (c,d) do respond to the surface tension.
In comparing our results with \citet{Giavedoni2007}, we used a surface tension of $70 \times 10^{-3}$~N/m where increasing surface elasticity and surface viscosity would preferentially increase the onset acceleration; however upon decreasing the surface tension, we find another crossover where increasing these surface parameters will decrease the onset acceleration.

For completeness, we have included figure \ref{f:avsRho} which graphs the onset acceleration $\mathcal{F}_{c}$ vs bulk fluid density $\rho$.
The onset acceleration generally decreases as bulk density increases.

\section{Conclusion and Discussion}
\label{s:Conclusion}

We have   derived an analytic expression for the onset acceleration for Faraday waves in a finite-depth infinite-breadth low-viscosity surfactant-covered fluid.
We have shown that this expression accurately reproduces the results of previous numerical works.
Our analysis required a novel definition of the Marangoni and Boussinesq numbers to handle the low-viscosity limit as the standard definition result in unbounded behavior for $\mathcal{F}_{c}$.

We have also shown that for shallow systems, the model model makes an unexpected prediction: adding a surfactant to a shallow system can lower the onset acceleration for Faraday waves.
In context of the energy-balance perspective of the emergence of Faraday waves, one would expect that by increasing the surface elasticity, which introduces a new viscous boundary layer at the free surface, one would increase the viscous dissipation and thereby increase the onset acceleration.
However, there are cases where increasing the bulk viscosity in the presence of a surfactant reduces the onset acceleration.
These unexpected results may be related to the work by \citet{Suman2008} where Marangoni effects in an inertial-less system act to destabilize the system.

We conclude by noting the potential utility of our analysis in determining the surface rheology of a surfactant monolayer.
Despite the myriad of surface rheometers that utilize macroscopic systems \citep{Fuller2012,Jaensson2018} and microscopic systems \citep{Samaniuk2014}, transverse and longitudinal capillary waves have long been used to probe the surface dilational viscosity \citep{Lemaire1992,Buzza1998,Saylor2000}, an historically difficult measurement.
Measuring the onset of Faraday waves is ideal for accessing the dilational viscosity since (i) no mechanical probe is introduced to the system's surface and (ii) at onset there are minimal surfactant concentration gradients, two key challenges that plague other measurement techniques \citep{Fuller2012}.
Further, the detection of Faraday waves requires a minimum of technical equipment.

Our analysis works effectively in finite-depth systems where the bulk fluid has a viscosity comparable to water (or less).
With our general solution, one could measure the onset acceleration for a range of frequencies and fit for the surface rheological parameters.
Marangoni and Boussinesq effects have different frequency dependencies and can be readily distinguished.
In fitting, one would determine $\Omega = \Lambda + 2 M$ rather than the surface dilational viscosity itself; however, with a surface shear viscometer, one could then deduce the dilational viscosity using Faraday waves.







\section{Acknowledgments}

We would also like to thank Lake Bookman for the many helpful discussions in the early attempts to formulate the theoretical framework.
We would also like to thank the NSF for grant \# NSF DMS-0604047 and NSF DMS-0968258.

Declaration of Interests: The authors report no conflict of interest.

\newpage

\appendix

\section{Integrating the Governing Equations}
\label{A1:Constants}

In \S \ref{s:GovernEqns}, we presented the governing equations for our model (eqns \ref{eqn:LinEqs}).
Here, we solve for $w$, $\zeta$, and $\Gamma$ at the moment that the Faraday waves emerge, the moment when these functions become non-trivial.
In this appendix, we will present an ansatz and solve for all but the final constants of integration, the wave mode amplitudes $\zeta_j$.
The analysis of these final constants of integration is addressed in \S\ref{s:Solving} and will yield an expression for the onset acceleration.

We use the following dimensionless ansatz:
\begin{equation*}
	\begin{split}
		w &= \cos( \vec{k} \cdot \vec{r}_H ) \sum_{j \in \mathbb{Z}_{\text{odd}}} i j w_{j}(z) e^{ i j t }  \\
		\zeta &= \cos( \vec{k} \cdot \vec{r}_H ) \sum_{j \in \mathbb{Z}_{\text{odd}}} \zeta_j e^{ i j t }  \\
		\Gamma &= 1 + \cos( \vec{k} \cdot \vec{r}_H ) \sum_{j \in \mathbb{Z}_{\text{odd}}} \Gamma_j e^{ i j t } 
	\end{split}
\end{equation*}
The dimensionless wave number $\vec{k}$ is not identically $1$ as the Faraday wavenumber is not equal to the wavenumber from the Kelvin dispersion relation.

With this ansatz, eqn \ref{eqn:LinEq-NSzModD} yields a family of 4th order linear ODEs for the $w_{j}(z)$ which can be readily solved.
	\begin{equation*}
		\left[ i j \mathcal{R}e \left( -k^2 + \partial_{zz} \right) - \left( -k^2 + \partial_{zz} \right)^2 \right] w_{j}(z) = 0
	\end{equation*}
	\begin{equation*}
		w_{j}(z) = \mathcal{A}_{j} \sinh( k z ) + \mathcal{B}_{j} \cosh( k z ) + \mathcal{C}_{j} \sinh( q_{j} z ) + \mathcal{D}_{j} \cosh( q_{j} z )
	\end{equation*}
where $q_{j}^2 = k^2 + i j \mathcal{R}e$.
The coefficients $\mathcal{A}_{j}$, $\mathcal{B}_{j}$, $\mathcal{C}_{j}$, and $\mathcal{D}_{j}$ are constants of integration which will be proportional to $\zeta_j$.

The surfactant continuity equation (eqn \ref{eqn:LinEq-Cont}) yields the coefficients of the surfactant distribution $\Gamma_j$.
\begin{equation*}
	\Gamma_j = \frac{ k \mathcal{A}_{j} + q_{j} \mathcal{C}_{j} }{ i j + \frac{k^2}{\mathcal{P}e} }
\end{equation*}

Eqns \ref{eqn:LinEq-BotBC1}, \ref{eqn:LinEq-Kin}, and \ref{eqn:LinEq-Tan} yield a system of equations for $\mathcal{A}_{j}$, $\mathcal{B}_{j}$, $\mathcal{C}_{j}$, and $\mathcal{D}_{j}$.
\begin{equation*}
	\begin{aligned}
		0 &= \mathcal{A}_{j} \sinh(-k H) + \mathcal{B}_{j} \cosh(-k H) + \mathcal{C}_{j} \sinh(-q_{j} H) + \mathcal{D}_{j} \cosh(-q_{j} H)
		\\
		0 &= k \mathcal{A}_{j} \cosh(-k H) + k \mathcal{B}_{j} \sinh(-k H) + q_{j} \mathcal{C}_{j} \cosh(-q_{j} H) + q_{j} \mathcal{D}_{j} \sinh(-q_{j} H)
		\\
		0 &= \zeta_j - \mathcal{B}_{j} - \mathcal{D}_{j} 
		\\
		0 &= \zeta_j k^2 + \left( k \mathcal{A}_{j} + q_{j} \mathcal{C}_{j} \right) \mathcal{S}_j + ( k^2 \mathcal{B}_{j} + q_{j}^2 \mathcal{D}_{j} ) 
	\end{aligned}
\end{equation*}

The solutions are:
\begin{equation*}
	\begin{aligned}
		\mathcal{A}_{j} &= \zeta_j \frac{ 
					\mathcal{S}_j q_{j} \mathcal{P}_{1j} 
					- 2 k^2 q_{j} \mathcal{P}_{3j}
					- (k^2 + q_{j}^2) \left( 
						k \mathcal{P}_{4j}
						- q_{j} 
					\right)
				}{ 
				\mathcal{Q}
				}
				\\
		\mathcal{B}_{j} &= - \zeta_j \frac{ 
					\mathcal{S}_j q_{j} \left(
						k \left( 1 - \mathcal{P}_{3j} \right)
						- q_{j} \mathcal{P}_{4j}
					\right)
					- (k^2 + q_{j}^2) \mathcal{P}_{2j}
				}{ 
				\mathcal{Q}
				}
				\\
		\mathcal{C}_{j} &= - \zeta_j \frac{ 
					\mathcal{S}_j k \mathcal{P}_{1j} 
					- 2 k^2 \left( k - q_{j} \mathcal{P}_{4j} \right)
					+ k (k^2 + q_{j}^2) \mathcal{P}_{3j}
				}{ 
				\mathcal{Q}
				}
				\\
		\mathcal{D}_{j} &= \zeta_j \frac{ 
					\mathcal{S}_j k \left(
						k \mathcal{P}_{4j}
						- q_{j} \left( 1 - \mathcal{P}_{3j} \right)
					\right)
					- 2 k^2 \mathcal{P}_{2j}
				}{ 
				\mathcal{Q}
				}
	\end{aligned}
\end{equation*}
where
\begin{equation*}
	\begin{aligned} 
		\mathcal{S}_j &= k^2 \left( \frac{ \mathcal{M} }{ i j + \frac{k^2}{\mathcal{P}e} } + \mathcal{B} \right) \\
		\mathcal{P}_{1j} &= q_{j} \tanh(H q_{j}) - k \tanh(H k) \\
		\mathcal{P}_{2j} &= q_{j} \tanh(H k) - k \tanh(H q_{j}) \\
		\mathcal{P}_{3j} &= \sech(H k) \sech(H q_{j}) \\
		\mathcal{P}_{4j} &= \tanh(H k) \tanh(H q_{j}) \\
		\mathcal{Q} &= - ( k^2 - q_{j}^2 ) \mathcal{P}_{2j} + \mathcal{S}_j \left( \left( k^2 + q_{j}^2 \right) \mathcal{P}_{4j} - 2 k q_{j} \left( 1 - \mathcal{P}_{3j} \right) \right)
	\end{aligned} 
\end{equation*}

In compiling all of these steps, we obtain the ansatz listed in eqn \ref{eqn:Ansatz} where the wave amplitude $\zeta_j$ is the only remaining unsolved constant of integration.

\section{Dominant Balance in the Weak-Viscosity Limit}
\label{A2:Expansion}

In \S\ref{s:Solving} we developed the central problem of this manuscript and mapped the route to the solution by considering a weak-viscosity fluid.
In expanding the pertinent quantities in terms of the expansion parameter $\Upsilon = \sqrt{\tfrac{1}{\mathcal{R}e}}$, we noted that that Marangoni and Boussinesq numbers had to be rescaled per the method of dominant balance.
Here, we detail the expansions for the pertinent quantities and their asymptotic behavior in the weak-viscosity limit.

The quantities $\mathcal{F}_{c}$, $k_{c}$, $q_{j}$, and $\mathcal{S}_j$  are expanded as:
\begin{equation*}
	\begin{aligned} 
		\mathcal{F}_{c} &= \sum_{n=1}^{\infty} \alpha_n \Upsilon^n 			
			\sim \mathcal{O} (\Upsilon) \\
		k_{c} &= 1 + \sum_{n=1}^{\infty} \beta_n \Upsilon^n 							
			\sim \mathcal{O} (1) \\
		q_{j}^2 &= k^2 + i j \tfrac{1}{\Upsilon^2} 														
			\sim \mathcal{O} (\tfrac{1}{\Upsilon^2}) \\
		\mathcal{S}_j &= \frac{k^2}{\Upsilon} \left( \frac{\mathcal{M}^{\dagger}}{ i j + \frac{k^2}{\mathcal{P}e} } + \mathcal{B}^{\dagger} \right) 				
			\sim \mathcal{O} (\tfrac{1}{\Upsilon})
	\end{aligned} 
\end{equation*}

In the low-viscosity limit,
\begin{equation*}
	\begin{aligned} 
		\tanh(H q_{j}) & \sim 1 + \mathcal{O}( e^{-\tfrac{H}{\Upsilon}} ) \rightarrow 1 \\
		\cosh(H q_{j}) & \sim \sinh(H q_{j}) \sim \mathcal{O}( e^{\tfrac{H}{\Upsilon}} ) \rightarrow \infty \\
		\sech(H q_{j}) & \sim \csch(H q_{j}) \sim \mathcal{O}( e^{- \tfrac{H}{\Upsilon}} ) \rightarrow 0
	\end{aligned} 
\end{equation*}
which simplifies the parameters $\mathcal{P}_{1j}$, $\mathcal{P}_{2j}$, $\mathcal{P}_{3j}$, and $\mathcal{P}_{4j}$:
\begin{equation*}
	\begin{aligned} 
		\mathcal{P}_{1j} & \rightarrow q_{j} - k \tanh(H k)  						
			\sim \mathcal{O} (\tfrac{1}{\Upsilon}) \\
		\mathcal{P}_{2j} & \rightarrow q_{j} \tanh(H k) - k 						
			\sim \mathcal{O} (\tfrac{1}{\Upsilon}) \\
		\mathcal{P}_{3j} & \rightarrow 0 											
			\sim \mathcal{O} (0) \\
		\mathcal{P}_{4j} & \rightarrow \tanh(H k) 														
			\sim \mathcal{O} (1) 
	\end{aligned} 
\end{equation*}

The constants of integration can then be expressed as:
\begin{equation*}
	\begin{aligned}
		\mathcal{Q} &= \mathcal{S}_j \left( \left( k^2 + q_{j}^2 \right) \mathcal{P}_{4j} - 2 k q_{j} \right) + \tfrac{i j}{\Upsilon^2} \mathcal{P}_{2j} 
			\sim \mathcal{O} (\tfrac{1}{\Upsilon^3}) \\
		\frac{\mathcal{A}_{j}}{\zeta_j} &= \frac{ 
					\mathcal{S}_j q_{j} + ( k^2 + q_{j}^2 )
				}{ 
				\mathcal{Q}
				} 
				\mathcal{P}_{1j} 
			\sim \mathcal{O} (1) \\
		\frac{\mathcal{B}_{j}}{\zeta_j} &= \frac{ 
					\mathcal{S}_j q_{j} + ( k^2 + q_{j}^2 )
				}{ 
				\mathcal{Q}
				}
				\mathcal{P}_{2j}
			\sim \mathcal{O} (1) \\
		\frac{\mathcal{C}_{j}}{\zeta_j} &= \frac{\mathcal{D}_{j}}{\zeta_j} = - \frac{ 
					\mathcal{S}_j k \mathcal{P}_{1j} 
					+ 2 k^2 \mathcal{P}_{2j} 
				}{ 
				\mathcal{Q}
				}
			\sim \mathcal{O} (\Upsilon) \\
	\end{aligned}
\end{equation*}

The coupling coefficient $H_j$ becomes:
\begin{equation*}
		H_j = - \frac{2}{G} \left[ 
					G
					+ \Sigma k^2 
					- \frac{j^2 \mathcal{S}_j q_{j} \mathcal{P}_{1j}}{k \mathcal{Q}}
					+ \frac{i j \Upsilon^2}{k \mathcal{Q}} \left(
						(k^2 + q_{j}^2)^2 \mathcal{P}_{1j}
						- 4 k^3 q_{j} \mathcal{P}_{2j}
						\right)
					\right]
\end{equation*}
In this form, one can easily check that $H_j \sim \mathcal{O}(1)$.
As mentioned in \S\ref{s:Solving}, the definitions of $\mathcal{M}^{\dagger}$ and $\mathcal{B}^{\dagger}$ ensure that the surfactant effects in the third term $\tfrac{j^2 \mathcal{S}_j q_j \mathcal{P}_{1j}}{k \mathcal{Q}}$ are $\mathcal{O}(1)$.

\section{Onset Acceleration}
\label{A3:SolnCoefs}

Here we present the expressions for the onset acceleration of all the cases given in \S\ref{s:Solutions}.
We also fully define all of the coefficients in the expressions.
These expressions were calculated with the aid of Mathematica.

\subsection{The surfactant-free infinite-depth limit}
\label{ass:InfDepthNoSurf}

\begin{equation}
	\label{Aeqn:CVAon}
	\mathcal{F}_{c} = \frac{ 1 }{ G } \left[ 8 \Upsilon^2 - 4 \sqrt{2} \Upsilon^3 + \frac{ 2 \sqrt{2} ( 11 - 2 G ) }{ ( 3 - 2 G ) } \Upsilon^5 + \mathcal{O}( \Upsilon^6 ) \right]
\end{equation}

\subsection{The surfactant-free finite-depth limit}
\label{ass:FinDepthNoSurf}

\begin{equation}
	\label{Aeqn:FDAon}
	\resizebox{.9\hsize}{!}{$
		\mathcal{F}_{c} = \frac{ 1 }{ G } \left[ 
			\sqrt{2} \csch^2(H)\Upsilon 
			+ \frac{ 4 \coth(H) \left(
			 4 \Sigma \cosh(2 H) + \cosh(3 H) \csch(H) + 4 H - 2 \Sigma
			 \right) }{ 
				\mathcal{Q}_{H}
			} \Upsilon^2
			+ \mathcal{O}( \Upsilon^3 ) 
		\right]
	$}
\end{equation}
where
\begin{align*}
	\mathcal{Q}_{H} &= 2 \Sigma \cosh(2 H) + \sinh(2 H) + 2 H - 2 \Sigma  \\
\end{align*}

\subsection{The infinite-depth surfactant limit}
\label{ass:InfDepthSurf}

\begin{equation}
	\label{Aeqn:AonSurfInfiniteH}
	\begin{aligned} 
		\mathcal{F}_{c} = \frac{ 1 }{ G } \left[ 
			\sqrt{2} \left( \frac{ 
				\mathcal{Q}_S
				- 1 
				+ \frac{ \sqrt{2} \mathcal{M}^{\dagger} }{ 1 + \tfrac{1}{\mathcal{P}e^2} } 
			}{
				\mathcal{Q}_S
			} \right) \Upsilon 
			+ \left( 
				\frac{ 
					2 \Sigma \mathcal{N}_1
					+ \mathcal{N}_2
				}{ 
					\left( 2 \Sigma + 1 \right) \left( 1 + \tfrac{1}{\mathcal{P}e^2} \right)^3 \left( 
					\mathcal{Q}_S
					\right)^3
				} 
			\right) \Upsilon^2 
			+ \mathcal{O}( \Upsilon^3 ) 
		\right]
	\end{aligned} 
\end{equation}
where the constants $\mathcal{Q}_S$, $\mathcal{N}_1$, and $\mathcal{N}_2$ are below:

\begin{subequations}
\begin{align*}
	\mathcal{Q}_S &= 
					1
					+ \sqrt{2} \mathcal{B}^{\dagger} 
					+ \mathcal{B}^{\dagger 2}
					+ \frac{ \mathcal{M}^{\dagger} }{ \mathcal{P}e \left( 1 + \tfrac{1}{\mathcal{P}e^2} \right) } \left( 
						\sqrt{2} 
						+ 2 \mathcal{B}^{\dagger} 
						- \sqrt{2} \mathcal{P}e
						+ \mathcal{M}^{\dagger} \mathcal{P}e
					\right)
\end{align*}
\begin{align*}
	\mathcal{N}_1 &= 
	\left( 1 + \tfrac{1}{\mathcal{P}e^2} \right)^3 \left( 8 + 20 \sqrt{2} \mathcal{B}^{\dagger} + 48 \mathcal{B}^{\dagger 2} + 34 \sqrt{2} \mathcal{B}^{\dagger 3} + 30 \mathcal{B}^{\dagger 4} + 8 \sqrt{2} \mathcal{B}^{\dagger 5} + 2 \mathcal{B}^{\dagger 6} \right) \\
	& \hspace{1em} + \left( 1 + \tfrac{1}{\mathcal{P}e^2} \right)^2 \mathcal{M}^{\dagger} \left( -20 \sqrt{2}  - 68 \mathcal{B}^{\dagger} - 54 \sqrt{2} \mathcal{B}^{\dagger 2} - 44 \mathcal{B}^{\dagger 3} - 8 \sqrt{2} \mathcal{B}^{\dagger 4} \right) \\
	& \hspace{1em} + \left( 1 + \tfrac{1}{\mathcal{P}e^2} \right)^2 \mathcal{M}^{\dagger 2} \left( 48 + 102 \sqrt{2} \mathcal{B}^{\dagger} + 180 \mathcal{B}^{\dagger 2} + 80 \sqrt{2} \mathcal{B}^{\dagger 3} + 30 \mathcal{B}^{\dagger 4} \right) \\
	& \hspace{1em} + \left( 1 + \tfrac{1}{\mathcal{P}e^2} \right) \mathcal{M}^{\dagger 2} \left( - 48 \sqrt{2} \mathcal{B}^{\dagger}  - 120 \mathcal{B}^{\dagger 2}  - 64 \sqrt{2} \mathcal{B}^{\dagger 3}  - 24 \mathcal{B}^{\dagger 4} \right) \\
	& \hspace{1em} + \left( 1 + \tfrac{1}{\mathcal{P}e^2} \right) \mathcal{M}^{\dagger 3} \left( -54 \sqrt{2} - 132 \mathcal{B}^{\dagger} - 48 \sqrt{2} \mathcal{B}^{\dagger 2} \right) 
		+ \left( 1 + \tfrac{1}{\mathcal{P}e^2} \right) \mathcal{M}^{\dagger 4} \left( 30 + 40 \sqrt{2} \mathcal{B}^{\dagger} + 30 \mathcal{B}^{\dagger 2} \right) \\
	& \hspace{1em} + \mathcal{M}^{\dagger 3} \left( 20 \sqrt{2} + 88 \mathcal{B}^{\dagger} + 32 \sqrt{2} \mathcal{B}^{\dagger 2} \right) 
		+ \mathcal{M}^{\dagger 4} \left( -32 \sqrt{2} \mathcal{B}^{\dagger} - 24 \mathcal{B}^{\dagger 2} \right) 
		+ \mathcal{M}^{\dagger 5} \left( -8 \sqrt{2} \right) 
		+ \mathcal{M}^{\dagger 6} \left( 2 \right) \\
	& \hspace{1em} + \left( 1 + \tfrac{1}{\mathcal{P}e^2} \right)^2 \mathcal{M}^{\dagger} \tfrac{1}{\mathcal{P}e} \left( +20 \sqrt{2} + 96 \mathcal{B}^{\dagger} + 102 \sqrt{2} \mathcal{B}^{\dagger 2} + 120 \mathcal{B}^{\dagger 3} + 40 \sqrt{2} \mathcal{B}^{\dagger 4} + 12 \mathcal{B}^{\dagger 5}\right) \\
	& \hspace{1em} + \left( 1 + \tfrac{1}{\mathcal{P}e^2} \right) \mathcal{M}^{\dagger 2} \tfrac{1}{\mathcal{P}e} \left( -68 - 108 \sqrt{2} \mathcal{B}^{\dagger}  - 132 \mathcal{B}^{\dagger 2} - 32 \sqrt{2} \mathcal{B}^{\dagger 3} \right) \\
	& \hspace{1em} + \left( 1 + \tfrac{1}{\mathcal{P}e^2} \right) \mathcal{M}^{\dagger 3} \tfrac{1}{\mathcal{P}e} \left( 34 \sqrt{2} + 120 \mathcal{B}^{\dagger} + 80 \sqrt{2} \mathcal{B}^{\dagger 2} + 40 \mathcal{B}^{\dagger 3} \right) \\
	& \hspace{1em} + \mathcal{M}^{\dagger 3} \tfrac{1}{\mathcal{P}e} \left(  20 \sqrt{2} - 32 \sqrt{2} \mathcal{B}^{\dagger 2} - 16 \mathcal{B}^{\dagger 3}  \right) 
		+ \mathcal{M}^{\dagger 4} \tfrac{1}{\mathcal{P}e} \left(  -44 - 32 \sqrt{2} \mathcal{B}^{\dagger} \right) 
		+ \mathcal{M}^{\dagger 5} \tfrac{1}{\mathcal{P}e} \left(  8 \sqrt{2} + 12 \mathcal{B}^{\dagger} \right) 
\end{align*}
\begin{align*}
	\mathcal{N}_2 &= \mathcal{N}_1 
									 + \left( 1 + \tfrac{1}{\mathcal{P}e^2} \right)^3 \mathcal{B}^{\dagger 3} \left( 2 \sqrt{2} + 4 \mathcal{B}^{\dagger} \right) \\
		& + \left( 1 + \tfrac{1}{\mathcal{P}e^2} \right)^2 \mathcal{B}^{\dagger} \mathcal{M}^{\dagger} \left( -4 + 2 \sqrt{2} \mathcal{M}^{\dagger} - 4 \sqrt{2} \mathcal{B}^{\dagger} + 12 \mathcal{B}^{\dagger} \mathcal{M}^{\dagger} - 8 \mathcal{B}^{\dagger 2} - 6 \sqrt{2} \mathcal{B}^{\dagger 3} \right) \\
		& + \left( 1 + \tfrac{1}{\mathcal{P}e^2} \right) \mathcal{B}^{\dagger} \mathcal{M}^{\dagger} \left( 16 \sqrt{2} \mathcal{M}^{\dagger} - 16 \mathcal{M}^{\dagger 2} + 24 \mathcal{B}^{\dagger} \mathcal{M}^{\dagger} - 24 \sqrt{2} \mathcal{B}^{\dagger} \mathcal{M}^{\dagger 2} + 8 \mathcal{B}^{\dagger 2} + 4 \sqrt{2} \mathcal{B}^{\dagger 3} \right) \\
		& + \mathcal{M}^{\dagger 2} \left( -4 \sqrt{2} \mathcal{M}^{\dagger} + 8 \mathcal{M}^{\dagger 2} - 2 \sqrt{2} \mathcal{M}^{\dagger 3} - 16 \sqrt{2} \mathcal{B}^{\dagger} + 16 \mathcal{B}^{\dagger} \mathcal{M}^{\dagger} - 24 \mathcal{B}^{\dagger 2} + 20 \sqrt{2} \mathcal{B}^{\dagger 2} \mathcal{M}^{\dagger} \right) \\
		& + \left( 1 + \tfrac{1}{\mathcal{P}e^2} \right)^2 \mathcal{B}^{\dagger} \mathcal{M}^{\dagger} \tfrac{1}{\mathcal{P}e} \left( 4 \sqrt{2} \mathcal{B}^{\dagger} + 12 \mathcal{B}^{\dagger 2} \right) \\
		& + \left( 1 + \tfrac{1}{\mathcal{P}e^2} \right) \mathcal{B}^{\dagger} \mathcal{M}^{\dagger} \tfrac{1}{\mathcal{P}e} \left( -4 \sqrt{2} \mathcal{M}^{\dagger} + 4 \mathcal{M}^{\dagger 2} + 4 \sqrt{2} \mathcal{B}^{\dagger} - 20 \mathcal{B}^{\dagger} \mathcal{M}^{\dagger} + 8 \mathcal{B}^{\dagger 2} - 20  \sqrt{2} \mathcal{B}^{\dagger 2} \mathcal{M}^{\dagger}\right) \\
		& + \mathcal{M}^{\dagger 2} \tfrac{1}{\mathcal{P}e} \left( -8 + 8 \sqrt{2} \mathcal{M}^{\dagger} - 4 \mathcal{M}^{\dagger 2} - 8 \sqrt{2} \mathcal{B}^{\dagger} + 32 \mathcal{B}^{\dagger} \mathcal{M}^{\dagger} - 12 \sqrt{2} \mathcal{B}^{\dagger} \mathcal{M}^{\dagger 2} + 16 \mathcal{B}^{\dagger 2} + 8 \sqrt{2} \mathcal{B}^{\dagger 3} \right) 
\end{align*}
\end{subequations}

\subsection{The general solution}
\label{ass:GenSoln}

\begin{equation}
	\label{Aeqn:AonGeneral}
	\resizebox{.9\hsize}{!}{$
	\begin{aligned} 
		\mathcal{F}_{c} =& \frac{ 1 }{ G } \left[ 
			\sqrt{2} \left( \frac{ 
				\mathcal{Q}_S \csch^2(H) 
				+ \left( \mathcal{Q}_S - 1 + \frac{ \sqrt{2} \mathcal{M}^{\dagger} }{ 1 + \tfrac{1}{\mathcal{P}e^2} } \right) \coth^2(H)
			}{
				\mathcal{Q}_S
			} \right) \Upsilon \right. \\
			& \hspace{2em} + \left( 
				\frac{ 
					\coth(H) \left(
						\cosh(2 H) \left( 4 \Sigma \mathcal{L}_1 + \mathcal{L}_2 \coth(H) \right)
						+ \cosh(3 H) \csch(H) \mathcal{L}_3
						+ 4 H \mathcal{L}_4
						- 2 \Sigma \mathcal{L}_5
						+ \coth(H) \mathcal{L}_6
					\right)
				}{ 
					\left( 1 + \tfrac{1}{\mathcal{P}e^2} \right)^3 \left( \mathcal{Q}_S \right)^3 \mathcal{Q}_H 
				} 
			\right) \Upsilon^2  \\
			& \left. \hspace{2em} + \mathcal{O}( \Upsilon^3 ) 
		\right]
	\end{aligned} 
	$}
\end{equation}
where
\begin{align*}
	\mathcal{L}_1 &= \tfrac{1}{2} \mathcal{N}_1
\end{align*}
\begin{align*}
	\mathcal{L}_2 &= \mathcal{N}_2 - 2 \mathcal{L}_3
\end{align*}
\begin{align*}
	\mathcal{L}_3 &= 
		\left( 1 + \tfrac{1}{\mathcal{P}e^2} \right)^3 \left( 4 + 17 \mathcal{B}^{\dagger 4} + \mathcal{B}^{\dagger 6} \right) \\
		& \hspace{1em} + \left( 1 + \tfrac{1}{\mathcal{P}e^2} \right)^2 \mathcal{M}^{\dagger} \left( -10 \sqrt{2} + 24 \mathcal{M}^{\dagger} + (10 \sqrt{2})/\mathcal{P}e -7 \sqrt{2} \mathcal{B}^{\dagger 4} + 15 \mathcal{M}^{\dagger} \mathcal{B}^{\dagger 4} + \tfrac{20 \sqrt{2}}{\mathcal{P}e} \mathcal{B}^{\dagger 4} \right) \\
		& \hspace{1em} + \left( 1 + \tfrac{1}{\mathcal{P}e^2} \right) \mathcal{M}^{\dagger} \left( 2 \sqrt{2} \mathcal{B}^{\dagger 4} - 12 \mathcal{M}^{\dagger} \mathcal{B}^{\dagger 4} - 27 \sqrt{2} \mathcal{M}^{\dagger 2} + 15 \mathcal{M}^{\dagger 3} + \tfrac{-34 \mathcal{M}^{\dagger} + 17 \sqrt{2} \mathcal{M}^{\dagger 2}}{\mathcal{P}e} \right) \\
		& \hspace{1em} + \mathcal{M}^{\dagger 2} \left( 8 \sqrt{2} \mathcal{M}^{\dagger} + 4 \mathcal{M}^{\dagger 2} - 5 \sqrt{2} \mathcal{M}^{\dagger 3} + \mathcal{M}^{\dagger 4} + \tfrac{ -4 + 14 \sqrt{2} \mathcal{M}^{\dagger} - 24 \mathcal{M}^{\dagger 2} + 4 \sqrt{2} \mathcal{M}^{\dagger 3} }{\mathcal{P}e} \right) 
\end{align*}
\begin{align*}
	\mathcal{L}_4 &= 
		\left( 1 + \tfrac{1}{\mathcal{P}e^2} \right)^3 \left(
			4 
			+ 10 \sqrt{2} \mathcal{B}^{\dagger} 
			+ 23 \mathcal{B}^{\dagger 2} 
			+ 14 \sqrt{2} \mathcal{B}^{\dagger 3} 
			+ 8 \mathcal{B}^{\dagger 4} 
			- \mathcal{B}^{\dagger 6} 
		\right) \\
		& \hspace{1em} + \left( 1 + \tfrac{1}{\mathcal{P}e^2} \right)^2 \mathcal{M}^{\dagger} \left(
			- 9 \sqrt{2} 
			+ 23 \mathcal{M}^{\dagger} 
			- 28 \mathcal{B}^{\dagger} 
			+ 42 \sqrt{2} \mathcal{M}^{\dagger} \mathcal{B}^{\dagger} 
			- 18 \sqrt{2} \mathcal{B}^{\dagger 2} 
			+ 48 \mathcal{M}^{\dagger} \mathcal{B}^{\dagger 2} 
			\right. \\
		& \hspace{10em} \left. 
			- 8 \mathcal{B}^{\dagger 3} 
			+ \sqrt{2} \mathcal{B}^{\dagger 4} 
			- 15 \mathcal{M}^{\dagger} \mathcal{B}^{\dagger 4} 
			+ \tfrac{10 \sqrt{2} + 46 \mathcal{B}^{\dagger} + 42 \sqrt{2} \mathcal{B}^{\dagger 2} + 32 \mathcal{B}^{\dagger 3} - 6 \mathcal{B}^{\dagger 5}}{\mathcal{P}e} 
		\right) \\
		& \hspace{1em} + \left( 1 + \tfrac{1}{\mathcal{P}e^2} \right) \mathcal{M}^{\dagger 2} \left(
			4 
			+ 18 \sqrt{2} \mathcal{M}^{\dagger} 
			- 8 \mathcal{M}^{\dagger 2} 
			+ 24 \sqrt{2} \mathcal{B}^{\dagger} 
			+ 24 \mathcal{B}^{\dagger} \mathcal{M}^{\dagger} 
			+ 36 \mathcal{B}^{\dagger 2} 
			\right. \\
		& \hspace{10em} \left. 
			- 6 \sqrt{2} \mathcal{B}^{\dagger 2} \mathcal{M}^{\dagger} 
			+ 15 \mathcal{B}^{\dagger 2} \mathcal{M}^{\dagger 2} 
			- 12 \mathcal{B}^{\dagger 4} 
			\right. \\
		& \hspace{10em} \left. 
			+ \tfrac{ 
				28
				-  14 \sqrt{2} \mathcal{M}^{\dagger}
				+ 36 \sqrt{2} \mathcal{B}^{\dagger}
				- 32 \mathcal{B}^{\dagger} \mathcal{M}^{\dagger}
				+ 24 \mathcal{B}^{\dagger 2}
				- 4 \sqrt{2} \mathcal{B}^{\dagger 3}
				+ 20 \mathcal{B}^{\dagger 3} \mathcal{M}^{\dagger}
				}{\mathcal{P}e}
		\right) \\
		& \hspace{1em} + \mathcal{M}^{\dagger 3} \left( 
			8 \sqrt{2} 
			- 4 \mathcal{M}^{\dagger} 
			+ \sqrt{2} \mathcal{M}^{\dagger 2} 
			- \mathcal{M}^{\dagger 3} 
			+ 16 \mathcal{B}^{\dagger} 
			- 4 \sqrt{2} \mathcal{B}^{\dagger 2} 
			\right. \\
		& \hspace{10em} \left. 
			+ 12 \mathcal{M}^{\dagger} \mathcal{B}^{\dagger 2} 
			+ \tfrac{ 4 \sqrt{2} 
			- 8 \mathcal{M}^{\dagger} 
			- 8 \mathcal{B}^{\dagger} 
			+ 4 \sqrt{2} \mathcal{M}^{\dagger} \mathcal{B}^{\dagger} 
			- 6 \mathcal{M}^{\dagger 2} \mathcal{B}^{\dagger} 
			+ 8 \mathcal{B}^{\dagger 3} }{\mathcal{P}e}
		\right) 
\end{align*}
\begin{align*}
	\mathcal{L}_5 &= 
		\left( 1 + \tfrac{1}{\mathcal{P}e^2} \right)^3 \left(
			4 
			+ 4 \sqrt{2} \mathcal{B}^{\dagger} 
			- 12 \mathcal{B}^{\dagger 2} 
			- 34 \sqrt{2} \mathcal{B}^{\dagger 3} 
			- 66 \mathcal{B}^{\dagger 4} 
			- 32 \sqrt{2} \mathcal{B}^{\dagger 5} 
			- 14 \mathcal{B}^{\dagger 6}
		\right) \\
		& \hspace{1em} + \left( 1 + \tfrac{1}{\mathcal{P}e^2} \right)^2 \mathcal{M}^{\dagger} \left(
			- 4 \sqrt{2} 
			+ 20 \mathcal{B}^{\dagger} 
			+ 54 \sqrt{2} \mathcal{B}^{\dagger 2} 
			+ 92 \mathcal{B}^{\dagger 3} 
			+ 32 \sqrt{2} \mathcal{B}^{\dagger 4} 
			- 12 \mathcal{M}^{\dagger} 
			- 102 \sqrt{2} \mathcal{B}^{\dagger} \mathcal{M}^{\dagger} 
			\right. \\
		& \hspace{10em} \left. 
			- 396 \mathcal{B}^{\dagger 2} \mathcal{M}^{\dagger} 
			- 320 \sqrt{2} \mathcal{B}^{\dagger 3} \mathcal{M}^{\dagger} 
			- 210 \mathcal{B}^{\dagger 4} \mathcal{M}^{\dagger} 
			\right. \\
		& \hspace{10em} \left. 
			+ \tfrac{
				4 \sqrt{2}
				- 24 \mathcal{B}^{\dagger}
				- 102 \sqrt{2} \mathcal{B}^{\dagger 2}
				-  264 \mathcal{B}^{\dagger 3}
				- 160 \sqrt{2} \mathcal{B}^{\dagger 4}
				- 84 \mathcal{B}^{\dagger 5}
			}{\mathcal{P}e}
		\right) \\
		& \hspace{1em} + \left( 1 + \tfrac{1}{\mathcal{P}e^2} \right) \mathcal{M}^{\dagger 2} \left(
			- 48 \sqrt{2} \mathcal{B}^{\dagger} 
			- 264 \mathcal{B}^{\dagger 2} 
			- 256 \sqrt{2} \mathcal{B}^{\dagger 3} 
			- 168 \mathcal{B}^{\dagger 4} 
			- 54 \sqrt{2} \mathcal{M}^{\dagger} 
			- 276 \mathcal{B}^{\dagger} \mathcal{M}^{\dagger} 
			\right. \\
		& \hspace{10em} \left. 
			- 192 \sqrt{2} \mathcal{B}^{\dagger 2} \mathcal{M}^{\dagger} 
			+ 66 \mathcal{M}^{\dagger 2} 
			+ 160 \sqrt{2} \mathcal{B}^{\dagger} \mathcal{M}^{\dagger 2} 
			+ 210 \mathcal{B}^{\dagger 2} \mathcal{M}^{\dagger 2} 
			\right. \\
		& \hspace{10em} \left. 
			+ \tfrac{
				- 20
				- 108 \sqrt{2} \mathcal{B}^{\dagger}
				- 276 \mathcal{B}^{\dagger 2}
				- 128 \sqrt{2} \mathcal{B}^{\dagger 3}
				+ 34 \sqrt{2} \mathcal{M}^{\dagger}
				+ 264 \mathcal{B}^{\dagger} \mathcal{M}^{\dagger}
				+ 320 \sqrt{2} \mathcal{B}^{\dagger 2} \mathcal{M}^{\dagger}
				+ 280 \mathcal{B}^{\dagger 3} \mathcal{M}^{\dagger}
			}{\mathcal{P}e}
		\right) \\
		& \hspace{1em} + \mathcal{M}^{\dagger 3} \left( 
			- 20 \sqrt{2} 
			- 184 \mathcal{B}^{\dagger} 
			- 128 \sqrt{2} \mathcal{B}^{\dagger 2} 
			+ 128 \sqrt{2} \mathcal{B}^{\dagger} \mathcal{M}^{\dagger} 
			+ 168 \mathcal{B}^{\dagger 2} \mathcal{M}^{\dagger} 
			+ 32 \sqrt{2} \mathcal{M}^{\dagger 2} 
			- 14 \mathcal{M}^{\dagger 3} 
			\right. \\
		& \hspace{10em} \left. 
			+ \tfrac{
				- 20 \sqrt{2}
				+ 128 \sqrt{2} \mathcal{B}^{\dagger 2}
				+ 112 \mathcal{B}^{\dagger 3}
				+ 92 \mathcal{M}^{\dagger}
				+ 128 \sqrt{2} \mathcal{B}^{\dagger} \mathcal{M}^{\dagger}
				- 32 \sqrt{2} \mathcal{M}^{\dagger 2}
				- 84 \mathcal{B}^{\dagger} \mathcal{M}^{\dagger 2}
			}{\mathcal{P}e}
		\right) 
\end{align*}
\begin{align*}
	\mathcal{L}_6 &= 
		\left( 1 + \tfrac{1}{\mathcal{P}e^2} \right)^3 \left(
			28 \mathcal{B}^{\dagger 2} 
			+ 48 \sqrt{2} \mathcal{B}^{\dagger 3} 
			+ 95 \mathcal{B}^{\dagger 4} 
			+ 32 \sqrt{2} \mathcal{B}^{\dagger 5} 
			+ 15 \mathcal{B}^{\dagger 6}
		\right) \\
		& \hspace{1em} + \left( 1 + \tfrac{1}{\mathcal{P}e^2} \right)^2 \mathcal{M}^{\dagger} \left(
			- 6 \sqrt{2} 
			- 48 \mathcal{B}^{\dagger} 
			- 94 \sqrt{2} \mathcal{B}^{\dagger 2} 
			- 140 \mathcal{B}^{\dagger 3} 
			- 57 \sqrt{2} \mathcal{B}^{\dagger 4} 
			+ 36 \mathcal{M}^{\dagger} 
			+ 116 \sqrt{2} \mathcal{B}^{\dagger} \mathcal{M}^{\dagger} 
			\right. \\
		& \hspace{10em} \left. 
			+ 432 \mathcal{B}^{\dagger 2} \mathcal{M}^{\dagger} 
			+ 320 \sqrt{2} \mathcal{B}^{\dagger 3} \mathcal{M}^{\dagger} 
			+ 225 \mathcal{B}^{\dagger 4} \mathcal{M}^{\dagger} 
			\right. \\
		& \hspace{10em} \left. 
			+ \tfrac{
				+ 6 \sqrt{2}
				+ 40 \mathcal{B}^{\dagger}
				+ 130 \sqrt{2} \mathcal{B}^{\dagger 2}
				+ 300 \mathcal{B}^{\dagger 3}
				+ 180 \sqrt{2} \mathcal{B}^{\dagger 4}
				+ 84 \mathcal{B}^{\dagger 5}
			}{\mathcal{P}e}
		\right) \\
		& \hspace{1em} + \left( 1 + \tfrac{1}{\mathcal{P}e^2} \right) \mathcal{M}^{\dagger} \left(
			16 \mathcal{B}^{\dagger} 
			+ 24 \sqrt{2} \mathcal{B}^{\dagger 2} 
			+ 40 \mathcal{B}^{\dagger 3} 
			+ 14 \sqrt{2} \mathcal{B}^{\dagger 4} 
			+ 24 \mathcal{M}^{\dagger} 
			+ 16 \sqrt{2} \mathcal{B}^{\dagger} \mathcal{M}^{\dagger} 
			- 192 \mathcal{B}^{\dagger 2} \mathcal{M}^{\dagger} 
			\right. \\
		& \hspace{10em} \left. 
			- 256 \sqrt{2} \mathcal{B}^{\dagger 3} \mathcal{M}^{\dagger} 
			- 180 \mathcal{B}^{\dagger 4} \mathcal{M}^{\dagger} 
			- 93 \sqrt{2} \mathcal{M}^{\dagger 2} 
			- 364 \mathcal{B}^{\dagger} \mathcal{M}^{\dagger 2} 
			\right. \\
		& \hspace{10em} \left. 
			- 264 \sqrt{2} \mathcal{B}^{\dagger 2} \mathcal{M}^{\dagger 2} 
			+ 81 \mathcal{M}^{\dagger 3} 
			+ 160 \sqrt{2} \mathcal{B}^{\dagger} \mathcal{M}^{\dagger 3} 
			+ 210 \mathcal{B}^{\dagger 2} \mathcal{M}^{\dagger 3} 
			\right. \\
		& \hspace{10em} \left. 
			+ \tfrac{
				+ 8 \sqrt{2}
				+ 32 \mathcal{B}^{\dagger}
				+ 28 \sqrt{2} \mathcal{B}^{\dagger 2}
				+ 24 \mathcal{B}^{\dagger 3}
				- 62 \mathcal{M}^{\dagger}
				- 160 \sqrt{2} \mathcal{B}^{\dagger} \mathcal{M}^{\dagger}
				- 392 \mathcal{B}^{\dagger 2} \mathcal{M}^{\dagger}
				- 188 \sqrt{2} \mathcal{B}^{\dagger 3} \mathcal{M}^{\dagger}
			}{\mathcal{P}e}
			\right. \\
		& \hspace{10em} \left. 
			+ \tfrac{
				+ 51 \sqrt{2} \mathcal{M}^{\dagger 2}
				+ 276 \mathcal{B}^{\dagger} \mathcal{M}^{\dagger 2}
				+ 320 \sqrt{2} \mathcal{B}^{\dagger 2} \mathcal{M}^{\dagger 2}
				+ 280 \mathcal{B}^{\dagger 3} \mathcal{M}^{\dagger 2}
			}{\mathcal{P}e}
		\right) \\
		& \hspace{1em} + \mathcal{M}^{\dagger 2} \left( 
			- 16 
			- 64 \sqrt{2} \mathcal{B}^{\dagger} 
			- 72 \mathcal{B}^{\dagger 2} 
			+ 24 \sqrt{2} \mathcal{M}^{\dagger} 
			+ 264 \mathcal{B}^{\dagger} \mathcal{M}^{\dagger} 
			+ 188 \sqrt{2} \mathcal{B}^{\dagger 2} \mathcal{M}^{\dagger} 
			+ 28 \mathcal{M}^{\dagger 2} 
			\right. \\
		& \hspace{10em} \left. 
			- 128 \sqrt{2} \mathcal{B}^{\dagger} \mathcal{M}^{\dagger 2} 
			- 168 \mathcal{B}^{\dagger 2} \mathcal{M}^{\dagger 2} 
			- 43 \sqrt{2} \mathcal{M}^{\dagger 3} 
			+ 15 \mathcal{M}^{\dagger 4} 
			\right. \\
		& \hspace{10em} \left. 
			+ \tfrac{
				- 28
				- 8 \sqrt{2} \mathcal{B}^{\dagger}
				+ 64 \mathcal{B}^{\dagger 2}
				+ 24 \sqrt{2} \mathcal{B}^{\dagger 3}
				+ 66 \sqrt{2} \mathcal{M}^{\dagger}
				+ 96 \mathcal{B}^{\dagger} \mathcal{M}^{\dagger}
				- 128 \sqrt{2} \mathcal{B}^{\dagger 2} \mathcal{M}^{\dagger}
			}{\mathcal{P}e}
			\right. \\
		& \hspace{10em} \left. 
			+ \tfrac{
				- 112 \mathcal{B}^{\dagger 3} \mathcal{M}^{\dagger}
				- 136 \mathcal{M}^{\dagger 2}
				- 164 \sqrt{2} \mathcal{B}^{\dagger} \mathcal{M}^{\dagger 2}
				+ 36 \sqrt{2} \mathcal{M}^{\dagger 3}
				+ 84 \mathcal{B}^{\dagger} \mathcal{M}^{\dagger 3}
			}{\mathcal{P}e}
		\right) 
\end{align*}

\bibliographystyle{unsrtnat}
\bibliography{Strickland}

\end{document}